\documentclass[a4paper]{article}
\usepackage{graphics,graphicx,array,hhline,color,float,mdwlist,bm,amsmath,amssymb,multirow,xcolor,tabularx,placeins,xspace}
\usepackage[left=2cm,right=2cm,top=2cm,bottom=2cm,head=1cm,foot=1cm]{geometry}
\usepackage[utf8]{inputenc}
\usepackage[magyar,english]{babel}
\usepackage{comment} 
\usepackage{caption}
\usepackage{booktabs}
\usepackage[hidelinks]{hyperref}
\usepackage[left]{lineno}
\usepackage{multicol}
\usepackage[style=phys,sorting=none,backend=biber,eprint,maxnames=3]{biblatex}
\graphicspath{{./figs/}}

\usepackage{graphicx} 
\usepackage{array}    

\newcolumntype{R}[1]{>{\centering\arraybackslash}p{#1}}

\author{The STAR Collaboration}
                  
\title{Measurement of pseudorapidity distributions from Au+Au collisions with the STAR Event Plane Detector}
\date{\today}

\begin{document}

\begin{titlepage}
\maketitle
\begin{abstract}
The measurement of charged-particle pseudorapidity distributions using the STAR Event Plane Detector (EPD) in gold-gold collisions at center-of-mass energies per nucleon pair of $\sqrt{s_{_{\text{NN}}}}=$ 200, 54.4, 27, 19.6, 14.6, 11.5, 9.2, and 7.7~GeV is presented. Charged particles are detected by the EPD within the $2.14<|\eta|<5.09$ pseudorapidity interval. The number of charged particles passing through each EPD ring is extracted using convoluted Landau fits to the deposited energy spectra. To account for the detector's response to the primary particles, these measured EPD hit numbers are unfolded using response matrices determined via Monte Carlo simulations. The response matrix is calculated separately for each energy and centrality class using the UrQMD event generator embedded into the GEANT3 simulation for the STAR detector. The new STAR measurements are consistent with the previous PHOBOS results in the corresponding rapidity intervals at 19.6 and 200 GeV, while providing improved precision. In addition, the new data extend the energy coverage to 7.7–200 GeV across multiple centrality classes. This expanded dataset enables a more comprehensive test of the limiting fragmentation hypothesis from central to peripheral collisions. The measurements are also compared with theoretical model calculations, including UrQMD, HIJING, and AMPT, in order to investigate the underlying physics implications.

\end{abstract}

\end{titlepage}

\begin{multicols}{2}
\raggedcolumns

\section{Introduction}

At extremely high temperatures and densities, strongly interacting matter undergoes a phase transition into a deconfined state of quarks and gluons known as the quark-gluon plasma (QGP)~\cite{STAR:2005gfr,ALICE:2022wpn}. This form of QCD matter is believed to have filled the early universe just microseconds after the Big Bang. Today, the QGP is recreated in ultra-relativistic collisions of heavy nuclei at facilities such as the Relativistic Heavy Ion Collider (RHIC) and the Large Hadron Collider (LHC), where its properties and the nature of the transition from confined to deconfined matter are actively studied~\cite{Busza:2018rrf,Chen:2024aom}.

The second phase of the Beam Energy Scan program (BES-II) at RHIC was conducted from 2018 to 2021, focusing on gold-gold (Au+Au) collisions. It covered a center-of-mass energy per nucleon pair range from $\sqrt{s_{_{\text{NN}}}}=$ 27 GeV to 7.7 GeV in collider mode, with the primary goal of exploring the QCD phase diagram in the high baryon density region~\cite{Yang:2017llt}. In preparation for BES-II, the Event Plane Detector (EPD)~\cite{Adams:2019fpo} was installed in 2017 at the Solenoidal Tracker at RHIC (STAR). The EPD provides an extended angular coverage  by detecting charged particles in the pseudorapidity range $2.14<|\eta|<5.09$, where pseudorapidity is defined as $\eta = -\ln[\tan(\theta/2)]$ in terms of the emission angle $\theta$ relative to the beam axis. The primary motivations for the EPD were to improve event plane resolution for flow measurements, provide an independent centrality determination for fluctuation measurements, and serve as a trigger detector during BES-II. The enhanced angular coverage, however, also allows charged-particle pseudorapidity ($\eta$) distributions to be measured over a wide forward/backward range.

The angular distribution of particles produced in heavy-ion collisions carries important information about the collision dynamics. In particular, the pseudorapidity distribution of charged particles, $dN_{ch}/d\eta$, is widely used to investigate the longitudinal dynamics of hot QCD medium evolution from the initial participating nucleons~\cite{Ke:2017squ,Shen:2017bsr,Jiang:2021bes}, and to tune and constrain theoretical models of heavy-ion collisions ~\cite{Huovinen:2025lct,Zhang:2024kdv,Schulz:2024imf}. Previous measurements~\cite{BRAHMS:2001llo, STAR:2001eyo, PHENIX:2000owy, PHOBOS:2010eyu} have typically focused on the midrapidity region and on collision energies above the BES-II energy range. For the success of the BES-II program, it is essential to measure $dN_{ch}/d\eta$ at the relevant energies and over a wide range in $\eta$, thereby providing a baseline for all theoretical models.

In this paper, we present the measurement of charged particle pseudorapidity distributions using the STAR EPD in Au+Au collisions at $\sqrt{s_{_{\text{NN}}}}=$ 200, 54.4, 27, 19.6, 14.6, 11.5, 9.2, and 7.7~GeV. The pseudorapidity distributions are extracted from EPD hit numbers via Bayesian iterative unfolding~\cite{unfold_theor} utilizing response matrices obtained from Monte Carlo (MC) simulations. The results are compared to previous measurements by the PHOBOS Collaboration~\cite{PHOBOS:2010eyu} and to theoretical models (UrQMD~\cite{urqmd1,urqmd2}, HIJING~\cite{hijing}, AMPT~\cite{ampt}). 

The paper is structured as follows: the experimental setup, with a focus on the EPD, is briefly described in Sec.~\ref{s:EPD}. The analysis procedure and systematic uncertainties are discussed in Sec.~\ref{s:analysis}. The results are presented and discussed in Sec.~\ref{s:results}, and a summary is provided in Sec.~\ref{s:summary}. Tabulated results are available in the HEPData entry for this analysis~\cite{hepdata}.

\section{Experimental setup} \label{s:EPD}

The general description of the STAR detector can be found in Ref.~\cite{STAR:2002eio}. Aside from event selection detectors, the only detector subsystem used in this analysis is the EPD~\cite{Adams:2019fpo}.

\begin{figure}[H]
\centering
\includegraphics[width=1\linewidth,trim={2cm 0 6cm 0},clip]{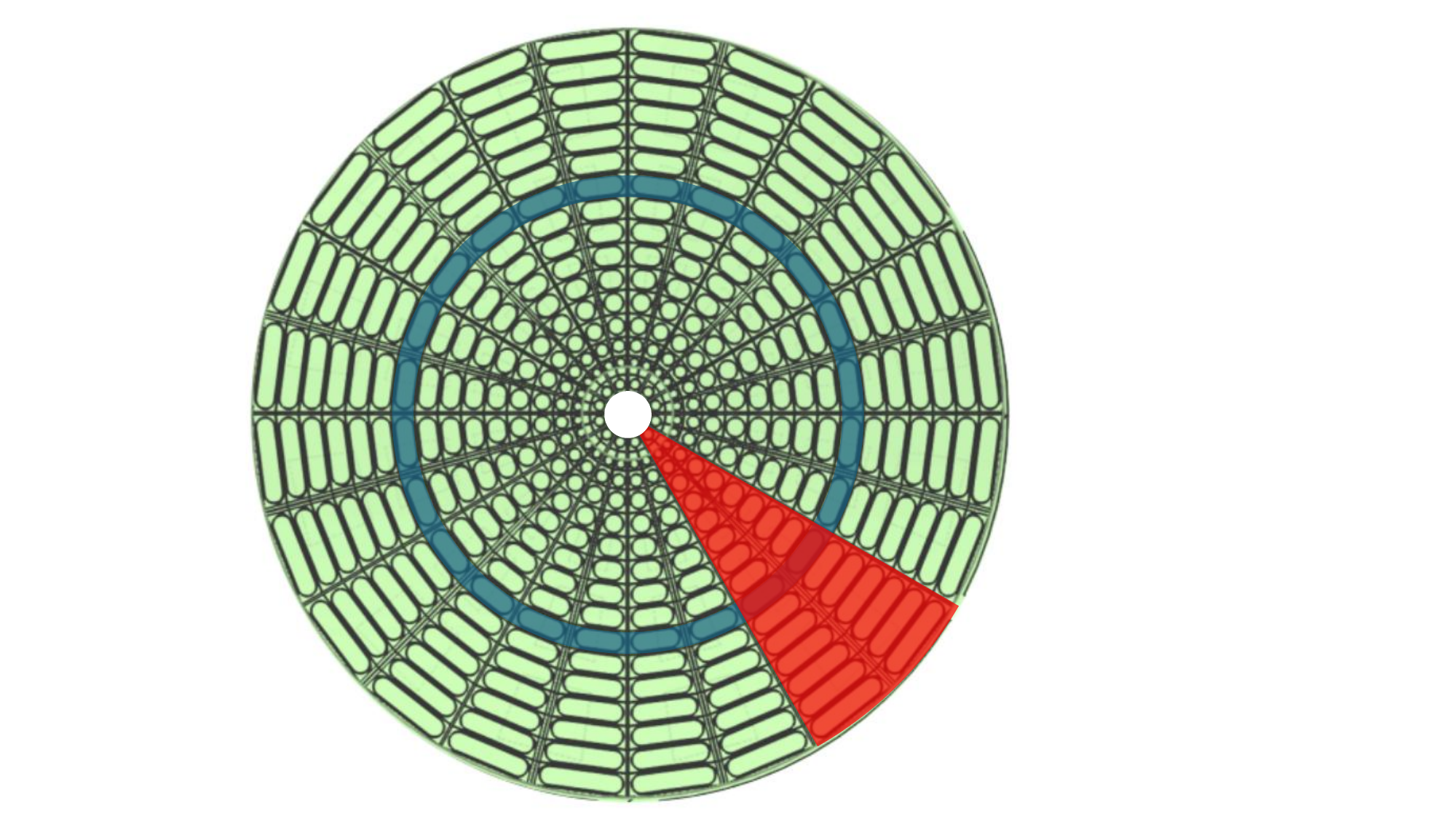}
\caption{Illustration of an EPD wheel. One of the 12 supersectors is highlighted in red, while one of the 16 rings is highlighted in blue.}
\label{f:epdstructure}
\end{figure}

The EPD comprises two highly segmented discs, or “wheels,” located at the East and West end of the STAR detector, each constructed with 1.2-cm-thick scintillator tiles.  Each wheel is divided into 12 “supersectors,” with each supersector covering an azimuthal angle of $30^{\circ}$ and further subdivided into 31 individual tiles. A schematic of a single wheel is shown in Fig.~\ref{f:epdstructure}. Wavelength-shifting fibers embedded in each tile collect and transport the light to fiber-to-fiber connectors, from which it is further transported via clear optical fibers to individual silicon photomultipliers (SiPMs) located on the STAR magnet poletips, some distance from the EPD. The resulting signals are then amplified and forwarded to the STAR digitizing and data acquisition system~\cite{Landgraf:2002zw}. The wheels are located at $\pm375$ cm from the nominal interaction point along the beam axis, oriented transverse to the beam. The only quantity measured by the EPD is the energy deposited by charged particles via ionization; as such, this detector cannot be used for particle identification. 

Instead of supersectors, an EPD wheel can also be considered as consisting of 16 rings, each covering a unique interval in $\eta$. Each ring contains 24 tiles, with the exception of the innermost ring on both sides, which has only 12 tiles. With this view, the relevant quantity for measuring $dN/d\eta$ is the index of the EPD ring. In this analysis, the rings on the East side are indexed from 0 to 15, moving outward from the center, while those on the West side are indexed from 16 to 31, moving inward from the outer edge. The minimum and maximum radial distance from the beam line with the corresponding minimum and maximum $\eta$ values are listed in Table~\ref{t:eta_multicol} for each EPD ring.

\begin{center}
\begin{tabular}{ccccc}
\toprule
\textbf{Ring} & $r_i$ (cm) & $r_f$ (cm) & $\eta_i$ & $\eta_f$ \\
\midrule
0  & 4.6    & 9.0    & 5.09 & 4.42 \\
1  & 9.0    & 13.4   & 4.42 & 4.03 \\
2  & 13.4   & 17.8   & 4.03 & 3.74 \\
3  & 17.8   & 23.3  & 3.74 & 3.47 \\
4  & 23.3  & 28.9  & 3.47 & 3.26 \\
5  & 28.9  & 34.4  & 3.26 & 3.08 \\
6  & 34.4  & 39.9  & 3.08 & 2.94 \\
7  & 39.9  & 45.5  & 2.94 & 2.81 \\
8  & 45.5  & 51.0  & 2.81 & 2.69 \\
9  & 51.0  & 56.5  & 2.69 & 2.59 \\
10 & 56.5  & 62.1  & 2.59 & 2.50 \\
11 & 62.1  & 67.6  & 2.50 & 2.41 \\
12 & 67.6  & 73.1  & 2.41 & 2.34 \\
13 & 73.1  & 78.6  & 2.34 & 2.27 \\
14 & 78.6  & 84.2  & 2.27 & 2.20 \\
15 & 84.2  & 89.7  & 2.20 & 2.14 \\
\bottomrule
\end{tabular}
\captionof{table}{Columns 2 and 3 contain the minimum ($r_i$) and maximum ($r_f$) radial distance from the beam line for each EPD ring, while columns 4 and 5 list the corresponding maximum ($\eta_i$) and minimum ($\eta_f$) pseudorapidity magnitudes for particles traveling in a straight line from the nominal interaction point~\cite{Adams:2019fpo}. The radial and pseudorapidity magnitudes apply to both EPD wheels, with opposite signs of $\eta$ for the East and West wheels.}
\label{t:eta_multicol}
\end{center}

\section{Measurement details} \label{s:analysis}

\subsection{Data sets}

The analysis was performed using minimum-bias data sets from Au+Au collisions at $\sqrt{s_{_{\text{NN}}}} = $ 200, 54.4, 27, 19.6, 14.6, 11.5, 9.2, and  7.7~GeV, recorded between 2017 and 2021 with the STAR detector. Minimum-bias events were selected by requiring a timing coincidence of signals on both sides of the interaction region in at least one of the following detectors: the Vertex Position Detector~\cite{Llope:2014nva}, the Zero Degree Calorimeters~\cite{Adler:2001fq}, or the inner rings of the EPD. Out-of-time pile-up events were removed by excluding outliers in the correlation between the number of Time Projection Chamber (TPC)~\cite{Anderson:2003ur} tracks and the number of those tracks matching a hit in the Time-of-Flight detector~\cite{Llope:2003ti}. Events were required to have a primary vertex, determined by TPC tracking, within 5 cm of the nominal interaction point along the z-axis. The transverse distance between the primary vertex and the beam line was required to be less than 1 cm.

The measurement was carried out in eight centrality classes: 0-5\%, 5-10\%, 10-20\%, 20-30\%, 30-40\%, 40-50\%, 50-60\%, and 60-70\%. Centrality determination was performed  based on charged-particle multiplicities measured by the TPC in the $|\eta| < 0.5$ region. These multiplicity distributions were corrected for luminosity and acceptance variation as a function of the z-vertex position ($V_z$), then fitted to MC Glauber simulations to determine centrality classes. Further details on the centrality determination can be found in Ref.~\cite{STAR:2012och}.

The tracking coverage of STAR does not overlap with the geometric acceptance of the EPD. After event selection, only the EPD signal was considered in this analysis, no traditional ``track selection'' was performed. 

The $\sqrt{s_{_{\text{NN}}}} = 54.4$~GeV data were collected during the first run with the EPD in 2017. At that time, only a quarter of a wheel was installed, and only on the East side. The measurement of $dN_{ch}/d\eta$ was still possible without too many additional complications, although only at negative $\eta$.

\subsection{Raw EPD data and Landau fits} \label{s:Landaufits}

The EPD measures the energy deposited by charged particles via ionization, which is recorded in units of Analog-to-Digital Converter (ADC) counts. Ionization is a probabilistic process; the deposited energy is described by a distribution, rather than a single value of the energy of the incoming particle. In the case of Minimum Ionizing Particles (MIPs) crossing a thin scintillator tile, as in the EPD, the deposited energy follows a Landau distribution~\cite{Adams:2019fpo}. In high-energy heavy-ion collisions, the vast majority of charged primary and secondary particles hitting the EPD in the forward region are MIPs or close to MIPs, and to first order, the Landau distribution provides an adequate description of the EPD tile response. The Landau distribution is characterized by two parameters: the most probable value (MPV) and the width of the distribution (WID). In practice, the ratio WID/MPV is often used instead of WID alone, as it depends only on the material and thickness of the detector.

In reality, multiple particles can strike a single EPD tile, and the goal is to extract the average number of particles from the observed ADC spectrum.  If $n$ MIPs hit a tile, the energy loss samples a probability distribution that is the convolution of $n$ Landau distributions. Since the number of particles hitting a tile varies from event to event, the overall ADC spectrum becomes a weighted sum of convoluted Landau distributions. To extract the average number of particles per tile, a convoluted Landau fit must be performed using the following fit function:
\begin{equation} \label{e:Landaufit}
    \frac{dN(x)}{d(\text{Calibrated ADC})}=\sum_{n=1}^{n_{\text{max}}}M_nL_n(x),
\end{equation} 
where each $M_n$ is a fit parameter (also called n-MIP weight) representing the probability of $n$ MIPs hitting the tile and $L_n$ are the convoluted Landau distributions:
\begin{equation}
    L_n(x)=\begin{cases} \text{Landau(x; MPV, WID)}, & \text{if} \hspace{0.2cm} n=1; \\
                     \int_{-\infty}^{\infty} L_{n-1}(x-y) L_1(y)dy, & \text{otherwise},
       \end{cases}
\end{equation}
where $x$ and $y$ are the variables of the distribution. 

Under realistic conditions, simple Landau distributions are sometimes unable to describe the data adequately. Particles, secondaries in particular, can hit the EPD at various angles, leading to a broadening of the spectra. Similarly, some particles may have lower momentum than a MIP and therefore deposit more energy, further contributing to the broadening. Finally, dark current and noise from the SiPMs is always present, which  further distorts the spectral shape. These effects can be taken into account by convoluting the Landau distribution with a Gaussian distribution. We found that the data are best described by performing the convolution over positive values only, i.e., using a one-sided Gaussian, leading to the following expression:
\begin{equation} \label{e:Langaufit}
    \frac{dN(x)}{d(\text{Calibrated ADC})}=\sum_{n=1}^{n_{\text{max}}}M_n\int_0^\infty  L_n(x-y)G(y)dy,
\end{equation} 
where $G(x)$ is a Gaussian distribution normalized over the interval $[0,\infty]$.

The fit including the Gaussian broadening has $n_{\text{max}}+3$ parameters: MPV, WID/MPV, Gaussian sigma, and the $M_n$ weights, where $n_{\text{max}}$ is the largest number of particles considered hitting the tile. For collision energies at $\sqrt{s_{_{\text{NN}}}}=27$~GeV or below, $n_{\text{max}}=4$ was sufficiently large. However, at $\sqrt{s_{_{\text{NN}}}}=54.4$~GeV and $\sqrt{s_{_{\text{NN}}}}=200$~GeV, values of $n_{\text{max}}$ up to 12 were required, depending on the centrality and ring position. The average number of particles hitting the tile ($N$) can be calculated as:
\begin{equation}
    N=\sum_{n=1}^{n_{\text{max}}}M_n\times n.
\end{equation}


Since we have a large amount of data, and the Landau shape of the ADC spectrum comes from a simple model, we do not expect a statistically perfect fit, i.e., a fit with $\chi^2/\text{NDF}\approx 1$, where NDF is the number of degrees of freedom. Analogously for a linear regression where the variance is estimated based on the residuals, and assuming that despite the large $\chi^2/\text{NDF}$ values the
$\chi^2$ contours are still valid, the uncertainties of the fit parameters obtained from the covariance matrix should be multiplied by $\sqrt{\chi^2/\text{NDF}}$~\cite{bevington2003data}. The uncertainty values shown in the figures already include this factor.

\begin{figure}[H]
\centering
\includegraphics[width=1\linewidth]{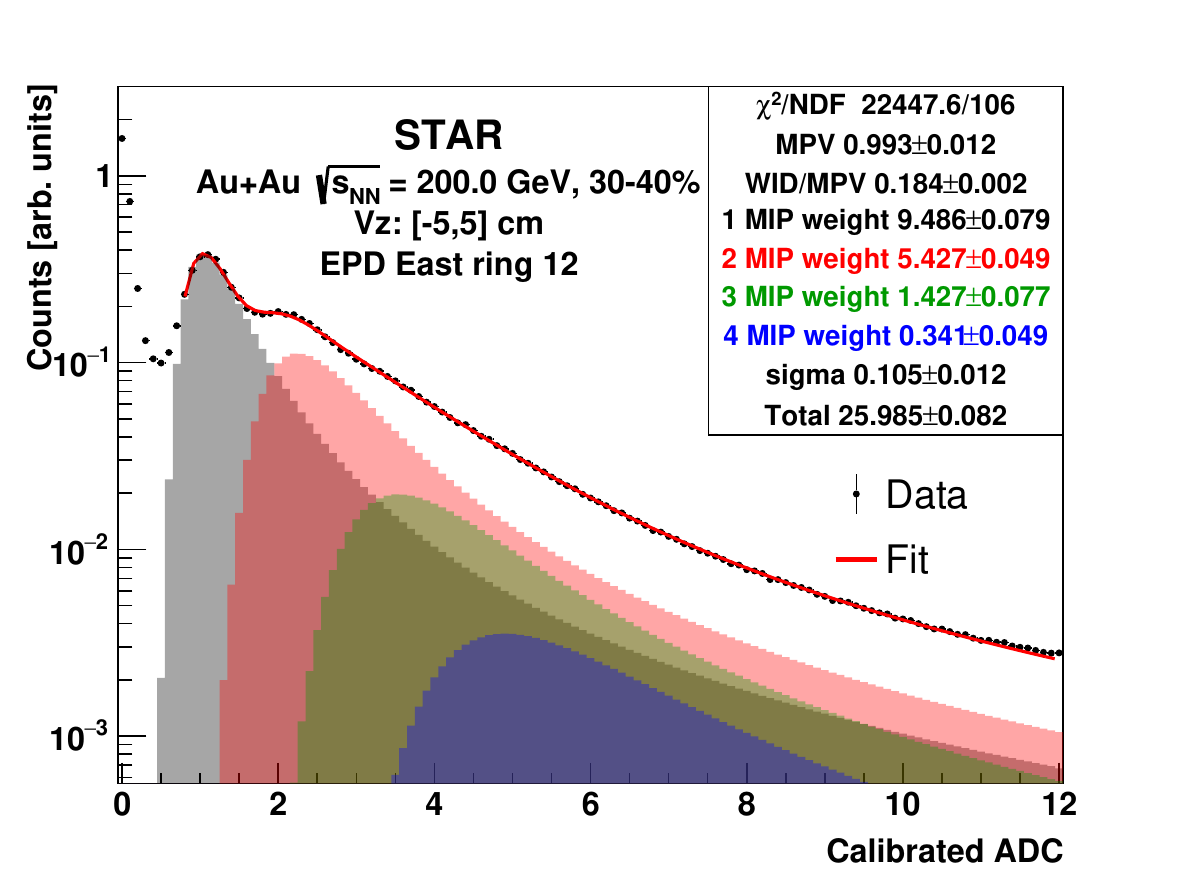}
\caption{Convoluted Landau fit to the calibrated ADC spectrum of a ring using the traditional fit method. The data is shown with black markers, while the red line represents the fit out to a calibrated ADC value of 12. The colored areas indicate the contributions from 1-, 2-, 3-, and 4-MIPs. The fit parameters are displayed in the legend. The average number of MIPs hitting the ring is shown as the ``Total''.}
\label{f:Landaufit}
\end{figure}

The ADC distributions measured for each individual EPD tile are calibrated such that the MPV corresponding to the single-MIP peak in each spectrum is normalized to unity. This calibration is done by first fitting the raw ADC spectra using convoluted Landau distributions, then scaling them by MPV to obtain calibrated ADC distributions. The Gaussian broadening is not included in the calibration fits, i.e., Eq.~\ref{e:Landaufit} is used.  In this analysis, we are interested only in the $\eta$ distribution, not the azimuthal distribution; therefore, the measurement is not performed with respect to the reaction plane, canceling most of the azimuthal dependence. (The azimuthal dependence of yields in the EPD and their systematics are discussed in Ref.~\cite{STAR:2024ujm}.) As a result, tiles within the same ring have nearly identical calibrated ADC distributions~\cite{Adams:2019fpo}, allowing the calibrated spectra from separate EPD tiles to be summed. While it is possible to perform the fits for each tile in a ring separately and then sum the average number of particles per tile, it is computationally more efficient to add the ADC spectra and perform a single fit for the entire ring. The fitting procedure follows the same approach described above. With proper normalization, the $M_n$ fit parameters correspond to the probability of $n$ MIPs hitting a single tile, multiplied by the number of tiles in the ring. Consequently, $N$ represents the average number of particles hitting the ring. An example fit for a ring is shown in Fig.~\ref{f:Landaufit}.

From this point onward, the fitting procedure described in the preceding part of this section  will be referred to as the ``traditional fit.'' It works well under normal circumstances; however, it was ultimately used only for systematics due to the complications discussed below.

\begin{figure}[H]
\centering
\includegraphics[width=1\linewidth]{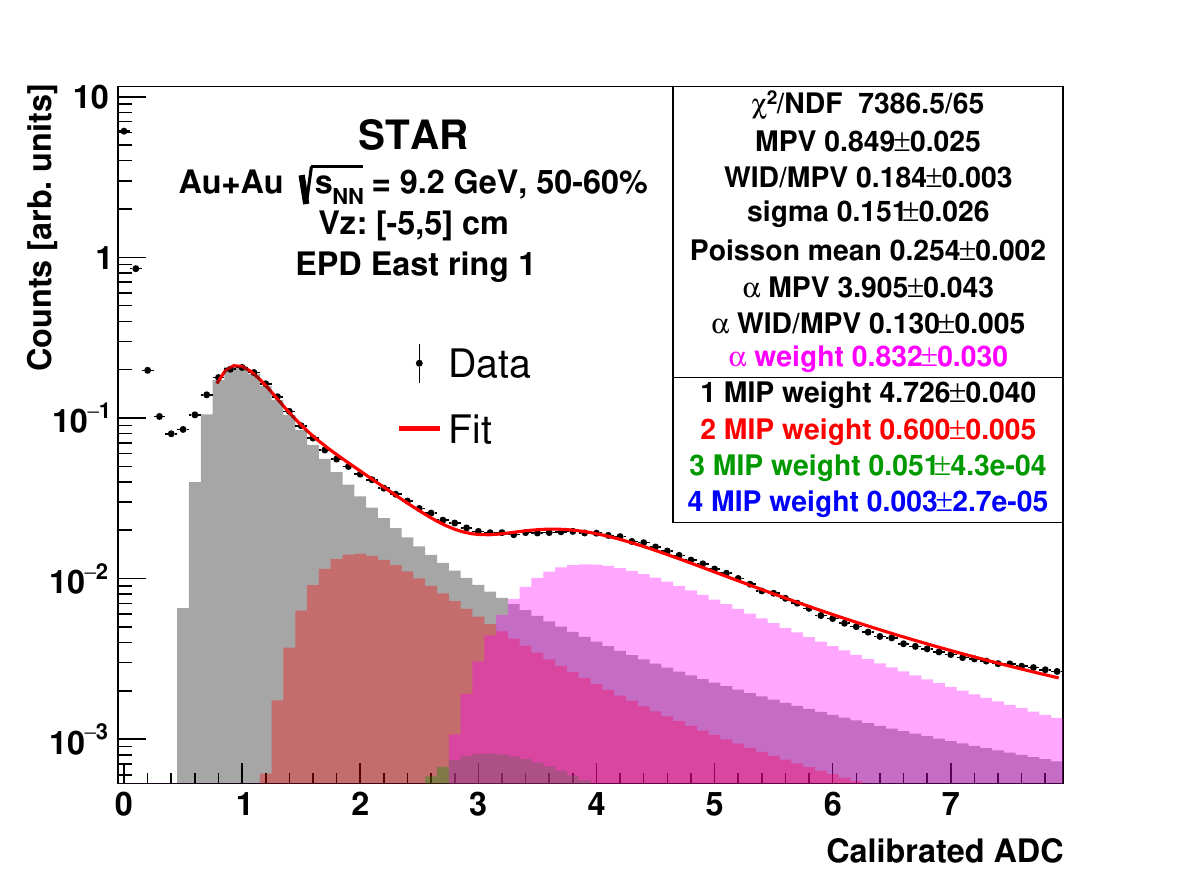}
\caption{Convoluted Landau fit to the calibrated ADC spectrum of a ring, where the contribution from alpha particles is evident, using the Poisson fit method with the alpha peak. The data is shown with black markers, while the red line represents the fit out to a calibrated ADC value of 8. The colored areas correspond to the 1-MIP, 2-MIP, 3-MIP, and $\alpha$ particle contributions, while the 4-MIP contribution is too small to be visible. The fit parameters are displayed in the legend (see text for detailed description). }
\label{f:Landaufitalphapeak}
\end{figure}

\begin{figure}[H]
\centering
\includegraphics[width=1\linewidth]{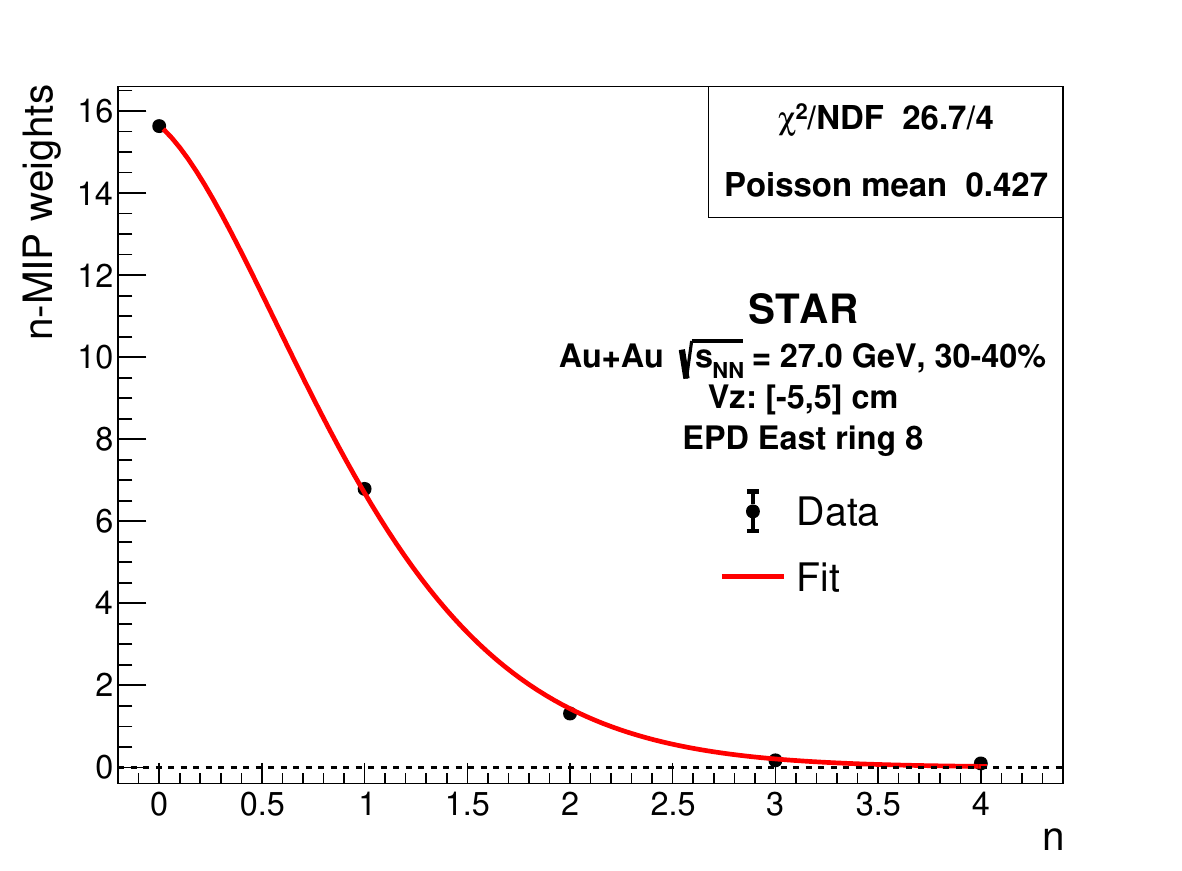}
\caption{n-MIP weights (black markers), obtained from a traditional Landau fit, fitted with a Poisson distribution (red line). The only fit parameter is the mean of the Poisson distribution. The $n=0$ point corresponds to tiles with zero hits.}
\label{f:Poissonfit}
\end{figure}

At lower collision energies, in peripheral centrality classes, we observed an extra contribution (a bump in the response spectrum) around 4 calibrated ADC in the spectra of the inner rings. An example of such a bump is shown in Fig.~\ref{f:Landaufitalphapeak}. This bump is unexpected from multiple singly charged MIPs because, due to the randomness of the process, the n-MIP weights should follow a Poisson distribution. This expectation is fulfilled in most cases, and is illustrated in Fig.~\ref{f:Poissonfit}, where the n-MIP weights obtained from a traditional Landau fit are fitted with a Poisson distribution. A large contribution (bump) around 4 calibrated ADC would imply a large 4-MIP weight, which violates the expectation from the Poisson distribution. Since this bump was absent in central collisions, disappeared at larger collision energies, and only appeared in the inner rings, we concluded that the bump is caused by particles with an atomic number ($Z$) of 2 (deposited energy is proportional to the square of the incoming particle's charge), originating from the projectile remnants, or in other words, spectators~\cite{Bialas:1976ed}. Although we cannot distinguish helium-3 from alpha particles, the abundance of alphas is far greater~\cite{Ogul:2010si}; therefore, we refer to and treat $Z=2$ particles as alphas. At the investigated energies, both protons and heavier fragments coming from spectators are relativistic and can be treated as MIPs. Unlike alpha particles, protons coming from spectators cannot be distinguished from singly-charged particles coming from the participant region.


In order to exclude alpha particles from the measurement, we added an additional term to the fit function in Eq.~\ref{e:Langaufit}, corresponding to the Landau distribution of alpha particles ($L_{\alpha}$):
\begin{align} \label{e:Landaufitalpha}
    \frac{dN(x)}{d(\text{Calibrated ADC})}=M_{\alpha}\int_0^\infty  L_{\alpha}(x-y)G(y)dy+ \notag \\
    +\sum_{n=1}^{n_{\text{max}}}M_n\int_0^\infty  L_n(x-y)G(y)dy, 
\end{align}
where $L_{\alpha}$ has its own independent parameters, MPV$_{\alpha}$ and WID$_{\alpha}$/MPV$_{\alpha}$, and its own weight $M_{\alpha}$. The probability of two alpha particles hitting the same tile is negligible, and therefore, this possibility is not included. These parameters are not used later in the analysis, apart from a correction discussed in Sec.~\ref{s:unfolding}.  The purpose of including $L_{\alpha}$ in the fit is solely to separate and exclude the contribution of alpha particles.  At $\sqrt{s_{_{\text{NN}}}} = 54.4$ and 200~GeV, there is no alpha contribution because spectators miss the EPD due to the larger collision energy. By excluding alphas (and all heavier fragments), we measure  $dN_{ch}/d\eta$ of singly-charged particles, including protons coming from spectators. Presenting a measurement of $dN_{ch}/d\eta$ of alpha particles and heavier fragments is possible, however, it requires slightly different methodology, therefore it is beyond the scope of this paper.

When performing the fit using the traditional method (including $L_{\alpha}$), we found that the 3- and 4-MIP weights were strongly correlated with the alpha weight, resulting in less reliable results. To address this, we developed a more robust fitting method that utilizes the Poisson distribution for the n-MIP weights. In this new procedure, we enforce the Poisson distribution of the n-MIP weights during the fit. This means that instead of having $n_{\text{max}}$ independent fit parameters, we now have only one parameter: the mean of the Poisson distribution. Since a Poisson distribution has only one parameter, all the n-MIP weights can be calculated from the Poisson mean. The other fit parameters (MPV, WID/MPV, sigma, $\alpha$ MPV, $\alpha$ WID/MPV, $\alpha$ weight) remained free and independent during the Poisson constrained fits. 

 The Poisson fitting method with $L_{\alpha}$ was used in all cases for consistency, but $M_{\alpha}$ was manually set to zero if no alpha peak was visible. A fit with an apparent alpha contribution is shown in Fig.~\ref{f:Landaufitalphapeak}. Although the individual MIP weights are shown, these are not fit parameters but are calculated from the Poisson mean. The average number of singly charged particles hitting the ring, calculated as the Poisson mean multiplied by the number of tiles in the ring, is shown as ``Total''. The relative uncertainty of the total yield is obtained from the fit as the relative uncertainty of the Poisson mean multiplied by $\sqrt{\chi ^2/\text{NDF}}$. Despite the strong correlation observed between results from the traditional and Poisson fitting methods, possible uncertainties arising from the choice of fitting method was accounted for in the systematic uncertainty.

At $\sqrt{s_{_{\text{NN}}}} = 27$~GeV and lower energies, the lower and upper limits of the fit were set to Calibrated ADC values of 0.75 and 8, respectively. The lower limit was chosen due to the presence of dark current below 0.75 calibrated ADC, and the upper limit was sufficiently large to encompass the largest considered n-MIP contribution (4-MIP). At $\sqrt{s_{_{\text{NN}}}} = 54.4$ and 200~GeV, the lower fit limit stayed the same, but larger n-MIP contributions had to be included, necessitating an increase in the upper fit limit. Depending on the number of n-MIP contributions included, the upper fit limit ranged from 8 to 24 calibrated ADC.  

At $\sqrt{s_{_{\text{NN}}}} = 200$~GeV, the innermost ring was saturated, i.e., so many MIPs hit each tile that the position of the 1-MIP peak could not be determined. Consequently, the innermost ring was not included in the analysis at this energy.

\subsection{Detector effects and response matrix} \label{s:responsematrix}

From the Landau fits, we obtained the average number of charged particles hitting each EPD ring in an event. Each ring corresponds to a unique interval in $\eta$ for a given vertex position, enabling the extraction of $dN_{ch}/d\eta$. In the following we detail this extraction procedure.

The main challenge is that we aim to measure $dN_{ch}/d\eta$ for primary particles, but the EPD is located several meters away from the primary vertex. Charged primary particles can scatter or decay in the detector material, producing secondaries that may also generate EPD hits. Even if a charged primary particle reaches the EPD, we cannot be certain that it hits the ring corresponding to its original $\eta$, as its momentum could have been altered while traversing the detector. Furthermore, neutral primaries can also contribute to EPD hits indirectly by producing charged secondaries, with the conversions occurring mainly in the beam pipe and support structures. The contribution from neutral primaries is non-negligible compared to that from charged ones. The largest contribution comes from $\pi^0\rightarrow\gamma\gamma$, with the photons converting into electron-positron pairs in the detector material.

Detector effects can be taken into account by introducing a response matrix. The linear relationship between the EPD hit count at each ring, $N(i_{\text{Ring}})$, and $dN/d\eta$ can be expressed as:
\begin{equation}
    N(i_{\text{Ring}})=\int R(\eta,i_{\text{Ring}})\frac{dN}{d\eta}d\eta,
\end{equation}
where $R(\eta,i_{\text{Ring}})$ is the response matrix that contains the number of hits in a given ring originating from a primary particle at a given $\eta$. Note that, for now, $dN/d\eta$ also includes neutral particles. The response matrix can be calculated using MC simulations. Once both the response and $N(i_{\text{Ring}})$ are known, the $dN/d\eta$ distribution can be obtained via unfolding. The unfolding procedure and the extraction of $dN_{\text{ch}}/d\eta$ are discussed in Sec.~\ref{s:unfolding}.

For practical purposes, the response matrix can be thought of as a two-dimensional histogram. In the simulation, we iterate over all primary particles. The number of particles per event and the total number of events are irrelevant, as long as the overall statistics are sufficient. For each primary particle, we construct a list of EPD hits originating from that primary. If no EPD hits are caused, either directly or indirectly by the primary, we record in the response matrix that the primary particle with pseudorapidity $\eta$ missed the EPD. If EPD hits are present, we iterate through them and fill the response matrix with the primary particle’s $\eta$ and the ring index of each hit. The response matrix obtained in this manner does not account for multiple hits from a single charged primary and does not correct for contributions from neutral primaries. These effects are discussed later in Sec.~\ref{s:unfolding}.

The calculation of the response matrix may seem straightforward, however, there are several subtleties that must be considered. When creating the list of EPD hits originating from a primary particle, hits caused by secondary particles created inside the EPD must be excluded. (When simulating the energy loss of a charged particle passing through a tile, GEANT will sometimes produce new particles such as electron--ion pairs, which then contribute to the energy loss. Naturally, we do not consider these as ``new particles'' which hit the tile, but only as details of how the energy of the original particle loses energy.) Additionally, EPD hits with very small deposited energy should also be excluded, specifically those below approximately 0.5 calibrated ADC. Even if such hits were present in the data, they would not be counted, as the Landau distribution falls off rapidly below this threshold. Lastly, if a single particle happens to create multiple hits in the EPD (without scattering or decaying outside between two hits), it still should be counted as a single hit. Of these three effects, the first impacts the final result by roughly 20\%, while the second and third combined contribute about a 5\% effect. Neglecting these effects would lead to an underestimation of the final $dN_{ch}/d\eta$.

An important consideration for the response matrix is the choice of event generator. Both HIJING~\cite{hijing} and UrQMD~\cite{urqmd1,urqmd2} were considered, but due to the importance of spectators, which are not included in HIJING, only UrQMD was utilized.

\begin{figure}[H]
\centering
\includegraphics[width=1\linewidth]{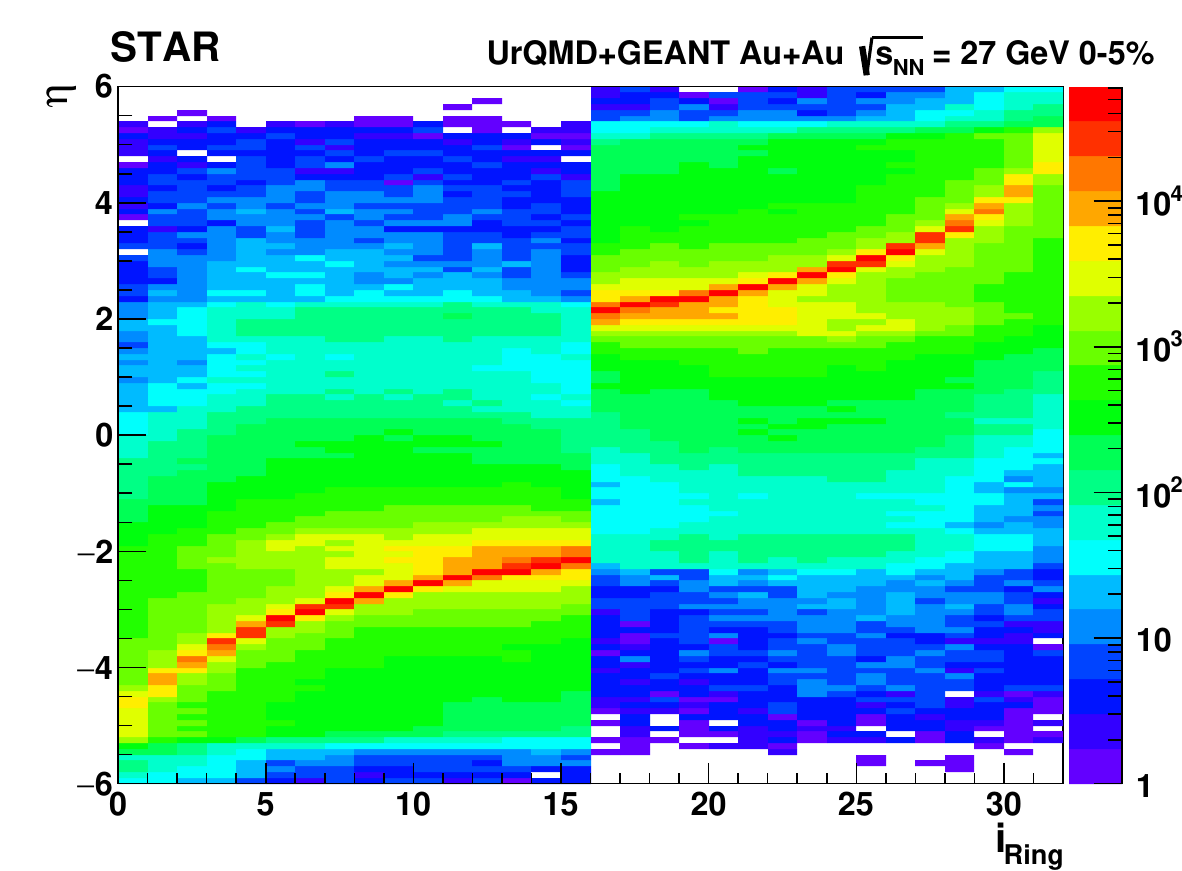}
\caption{Illustration of the response matrix calculated using UrQMD+GEANT3 at $\sqrt{s_{_{\text{NN}}}}= 27$~GeV in the 0-5\% centrality class. The left (right) side corresponds to the East (West) EPD wheel.}
\label{f:response}
\end{figure}

We also found that a single response matrix cannot be used for all energies and centralities. Several properties of the collision, such as $dN/d\eta$, the charged-to-neutral particle ratio, and the baryon-to-meson ratio vary with both energy and centrality, and each of these properties influences the response matrix. To achieve the most accurate results, a separate response matrix was generated for each collision energy and centrality class using MC simulations based on UrQMD events at the corresponding energy and centrality, with the STAR material budget simulated using GEANT3~\cite{geant}. A response matrix is shown in Fig.~\ref{f:response} for a selected energy and centrality class. The red ``diagonal'' areas correspond to the $\eta$ region covered by each ring. Due to secondaries, the response matrix is not fully diagonal; for example, the green areas at midrapidity arise from secondaries hitting the EPD that originate from a primary at midrapidity, while the blue areas correspond to backscattered secondaries. The yellow regions around $\left|\eta\right|=2$ are caused by the increased number of secondaries from TPC and beampipe support structures.

The response matrices obtained at $\sqrt{s_{_{\text{NN}}}} = 54.4$~GeV were somewhat different from those at other energies. Since only a quarter of an EPD wheel was present during data taking at $\sqrt{s_{_{\text{NN}}}} = 54.4$~GeV, and only on the East side, the remaining parts of the East EPD wheel as well as the entire West EPD wheel were ``deactivated'' in the simulations. In this way, no assumption of azimuthal uniformity was required, and no additional systematic uncertainties were introduced. Visually, the only difference compared to other energies is the empty ``right'' side of the matrices, corresponding to the missing West EPD wheel.

Note that the response matrix is not a differential distribution, thus its contents are not rescaled by bin size. To better match the widths of the $\eta$ regions covered by each EPD ring, we used non-uniform binning in $\eta$ (e.g., this binning does not correspond to the values listed in Table~\ref{t:eta_multicol}). Since non-uniform binning would distort the appearance of the response matrix, the illustration shown in Fig.~\ref{f:response} has uniform binning. The final $dN/d\eta$ distribution uses the same non-uniform $\eta$ binning as the response matrix. 


Of course, we cannot be certain that UrQMD realistically reproduces all of the relevant properties, such as $dN/d\eta$, the charged-to-neutral particle ratio, and the baryon-to-meson ratio. Therefore we re-ran the analysis with modified versions of these distributions to study the effects.  The resulting systematic uncertainties are discussed in Secs.~\ref{s:inputdNdeta} and~\ref{s:chargedneutral}.

\subsection{Extracting pseudorapidity distributions of singly-charged particles} \label{s:unfolding}

Before the extraction of $dN/d\eta$, a correction must be be applied to the $N(i_{\text{Ring}})$ values obtained from the Landau fits at energies where alpha particles from the spectator region reach the EPD. As they traverse material like the beam pipe, alpha particles can produce singly-charged secondary particles. These secondaries may also reach the EPD and are indistinguishable from other singly-charged particles. Consequently, the $N(i_{\text{Ring}})$ values include contributions from alpha-induced secondaries. Although UrQMD includes spectators, it does not produce heavier spectator fragments such as alpha particles. Therefore, to use the response matrices generated with UrQMD, the contribution from singly-charged secondaries produced by alpha particles must be subtracted from the $N(i_{\text{Ring}})$ values for each EPD ring. 

The number of alpha particles hitting each EPD ring is obtained from the Landau fits discussed in Sec.~\ref{s:Landaufits}. However, not all primary alphas reach the EPD, and even those that do, do not always produce a detectable signal. To account for this, an efficiency correction calculated using Geant4~\cite{Allison:2016lfl} is applied, since  Geant3 does not include hadronic interactions of alpha particles. An estimate of the primary alpha-particle distribution is then obtained by dividing the efficiency-corrected yields by the width of the $\eta$ interval corresponding to each EPD ring. 

To obtain the number of alpha-induced secondaries, alpha particles are generated according to their primary $\eta$ distribution discussed above, and the STAR detector response is simulated with Geant4. The momentum of each alpha particle is estimated by assuming that its magnitude equals that of an alpha particle traveling with the beam velocity of the incoming nucleus. The resulting numbers of singly-charged secondaries hitting each ring are subtracted from the $N(i_{\text{Ring}})$ values. In the most extreme cases, this correction can be as large as 30-40\%, but it decreases rapidly with increasing collision energy and outside the spectator $\eta$ region.

After obtaining the corrected $N(i_{\text{Ring}})$ values, the extraction of $dN/d\eta$ for singly-charged particles can begin.  In a simple scenario, one could obtain $dN/d\eta$ by inverting the response matrix. However, since the response matrix is not guaranteed to be invertible, straightforward matrix inversion may not be feasible. Even when the non-invertibility is only approximate due to uncertainties in the matrix, a straightforward numerical inversion can lead to large computational errors. Therefore, a more robust approach is to solve such equations using an unfolding method. This analysis employs Bayesian iterative unfolding, as described in Refs.~\cite{unfold_theor,unfold_theor_new}.

In practice, Bayesian iterative unfolding is performed using the \texttt{RooUnfold} package~\cite{roounfold_package,roounfold}. It is implemented in \texttt{C++}, built on \texttt{ROOT}~\cite{Brun:1997pa} classes, and runs within the \texttt{ROOT} environment. The input objects are the response matrix and the $N(i_{\text{Ring}})$ values. The unfolding uses the pseudorapidity distribution (encoded in the response matrix) produced by UrQMD as the prior, and returns a pseudorapidity distribution in the end. However, as discussed below, further corrections still need to be applied.

As mentioned earlier, a single primary particle can produce multiple EPD hits through secondaries, and neutral primaries can also induce EPD hits indirectly via charged secondaries. These effects are not automatically accounted for during the unfolding process using the response matrix described in Sec.~\ref{s:responsematrix}, so corrections must be applied afterward. From a mathematical standpoint, these corrections could also be incorporated directly into the response matrix, which would yield essentially the same final result, however, implementing the latter approach in practice is less straightforward.

\begin{figure}[H]
    \centering
    \includegraphics[width=1\linewidth]{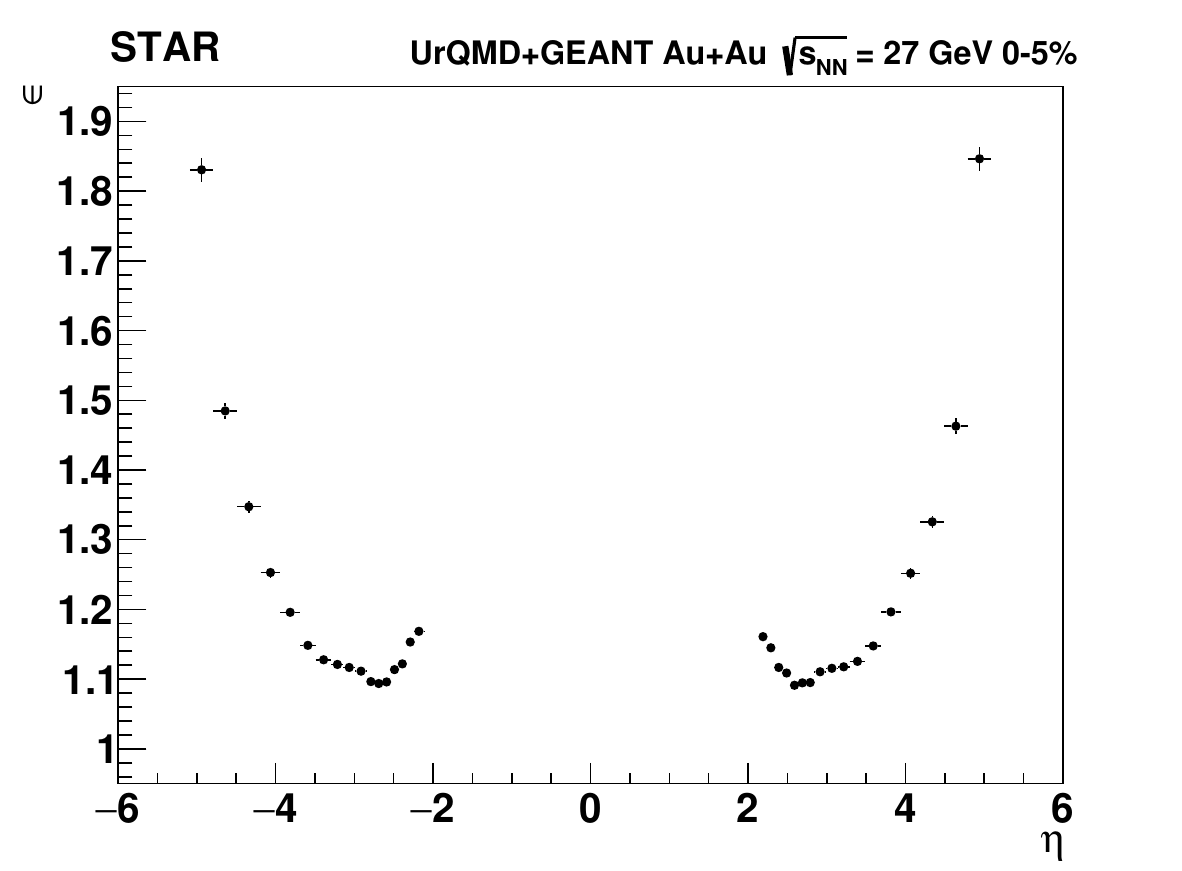}
    \caption{The average number of EPD hits ($\epsilon$) created by a primary particle  with pseudorapidity $\eta$ that caused at least 1 hit at $\sqrt{s_{_{\text{NN}}}}= 27$~GeV in the 0-5\% centrality class. The peaks around $|\eta|\approx2.2$ and $|\eta|\approx5$ are due to support structures.}
    \label{f:effcorr}
\end{figure}

To account for multiple EPD hits originating from the same primary particle, the average number of EPD hits ($\epsilon$) caused by a primary particle that is responsible for at least 1 hit, is calculated as a function of the primary particle's pseudorapidity using the same UrQMD+GEANT3 simulation employed for the response matrix. The unfolded $dN/d\eta$ distribution is then divided by this factor. The $\eta$ dependence of $\epsilon$ is shown in Fig.~\ref{f:effcorr} as an example for one energy and centrality.

\begin{figure}[H]
    \centering
    \includegraphics[width=1\linewidth]{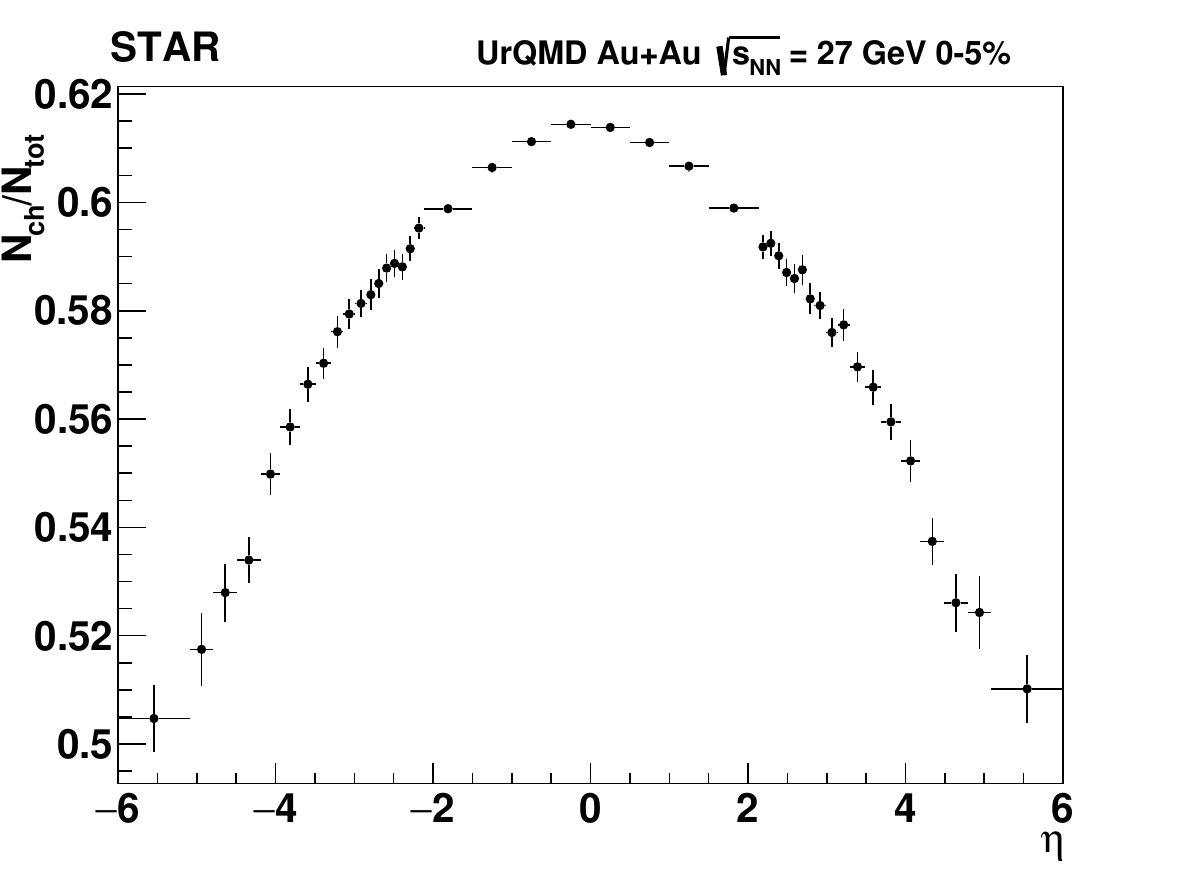}
    \caption{The fraction of charged primaries ($N_{\text{ch}}/N_{\text{tot}}$) as a function of $\eta$ at $\sqrt{s_{_{\text{NN}}}}= 27$~GeV in the 0-5\% centrality class.}
    \label{f:chargedcorr}
\end{figure}

Similarly, the fraction of charged primaries ($N_{\text{ch}}/N_{\text{tot}}$) is calculated as a function of $\eta$ using the same UrQMD simulation. The unfolded and multiple-counting-corrected $dN/d\eta$ is then multiplied by $N_{\text{ch}}/N_{\text{tot}}$ to obtain the physical pseudorapidity distribution of charged primaries, $dN_{\text{ch}}/d\eta$. A typical $\eta$ dependence of $N_{\text{ch}}/N_{\text{tot}}$ is shown in Fig.~\ref{f:chargedcorr} for the same energy and centrality as Fig.~\ref{f:effcorr}.

A closure test was performed by comparing the true MC $dN_{\text{ch}}/d\eta$ distribution with that obtained by unfolding the simulated EPD hit numbers and applying the corrections described above. The procedure showed agreement within 1\% between the true and reconstructed distributions.

\subsection{Systematic uncertainties} \label{s:syst}

Several systematic uncertainty sources were identified during the analysis. These sources and the method for calculating the total systematic uncertainty are presented in this section.

\subsubsection{Centrality selection} 

As mentioned previously, centrality selection is based on charged particle multiplicities measured by the TPC. To estimate the uncertainty from the entire centrality calibration procedure, the minimum and maximum multiplicity values were simultaneously varied by $\pm 5\%$ for each centrality class. The Landau fits were performed using the modified centrality classes, then the unfolding was done with the default response matrices, resulting in a systematic uncertainty of about 4\%.

\subsubsection{z-vertex resolution}

The final results are dependent on the z-vertex position of the included events; therefore, only events with $|V_z|<5$~cm were considered. To estimate the effects arising from the z-vertex resolution, this interval was shifted by $\pm 5$~cm, resulting in a systematic uncertainty of about 1\%. Since the goal of this study was to assess the effect of a $\pm 5$~cm offset in the reconstructed z-vertex position, the shift was applied only to the data and not to the MC sample used to construct the response matrix (see also the next item). This study was performed only at $\sqrt{s_{_{\text{NN}}}} = 27$~GeV. As the effect is primarily geometrical and relatively small, the resulting uncertainty was assumed to be independent of collision energy and was therefore applied to the results at all other collision energies.

\subsubsection{z-vertex choice} \label{s:zvertexchoice}

By default, only events with a primary z-vertex between -5 and 5 cm were used, and the same interval was applied in the MC simulation when calculating the response matrix. The MC events were generated with a uniform z-vertex distribution within this interval. Changing the z-vertex of events also alters the $\eta$ range corresponding to each EPD ring. If the same z-vertex position distribution is used in the MC as in the collision data, the unfolded $dN_{\text{ch}}/d\eta$ should, in principle, remain independent of the z-vertex selection choice. However, after analyzing multiple z-vertex intervals between -45 and 45 cm, we observed a systematic dependence on the z-vertex position. For events with $V_z=40$ cm, the EPD acceptance is shifted by approximately 0.1 units in $\eta$ relative to the nominal acceptance, which may explain the observed $z$-vertex dependence. Note, this study also provides a test of the robustness of the overall unfolding procedure, including the modeling of the detector material and support structures in the response matrix, under variations in the z-vertex position. 

To estimate the systematic uncertainty associated with the z-vertex choice, we considered two z-vertex intervals: $[-45, -35]$~cm and $[35, 45]$~cm. The same intervals were applied in the corresponding simulations. This study was performed at $\sqrt{s_{_{\text{NN}}}}= 27$ and 9 GeV. Since no significant dependence on collision energy was observed, the average 3\% uncertainty obtained at $\sqrt{s_{_{\text{NN}}}}= 27$~GeV was assumed to be collision-energy independent and applied to the results at all other energies. The average uncertainty for $\sqrt{s_{_{\text{NN}}}}= 200$~GeV shown in Table~\ref{t:syst} is smaller because data from the innermost ring is not included at that energy.

\subsubsection{Modifying the input MC $dN/d\eta$} \label{s:inputdNdeta}

The most important quantity in the MC simulation used to create the response matrix is $dN/d\eta$ of the particles present in the simulation. Even though Bayesian iterative unfolding is employed, the unfolded $dN_{\text{ch}}/d\eta$ still exhibits a slight dependence on the ``input'' MC $dN/d\eta$. We attempted to estimate the shape of $dN/d\eta$ as accurately as possible by using a realistic simulation at the same collision energy and centrality as the data. However, there is no guarantee that the $dN/d\eta$ produced by the simulation is truly realistic. To estimate the systematic uncertainty, the input $dN/d\eta$ was modified using different methods, and the unfolding was performed with the modified inputs.

When creating the response matrix, any desired input $dN/d\eta$ shape can be generated by ``reshaping'' the $dN/d\eta$ produced by UrQMD. This reshaping can be achieved by selectively rejecting particles according to an $\eta$-dependent factor. Three types of modifications were applied separately: the original $dN/d\eta$ was made wider or narrower; it was scaled to match the UrQMD shape at a higher or lower collision energy; it was shaped to match the unfolded $dN_{\text{ch}}/d\eta$. To obtain a wider shape, the probability of keeping a track ($p_{\text{keep}}$) was taken as:
\begin{equation}
    p_{\text{keep}}=e^{\frac{\eta^2-\eta_{\text{max}}^2}{2\sigma_{\text{broad}}^2}},
\end{equation}
with $\eta_{\text{max}}=6$ and $\sigma_{\text{broad}}=4$. A slightly different factor was used to obtain a narrower shape:
\begin{equation}
    p_{\text{keep}}=e^{-\frac{\eta^2}{2\sigma^2}},
\end{equation}
with  $\sigma=3$.

 The average uncertainty from widening/narrowing, shaping to higher/lower collision energy, and shaping to unfolded results was around 6\%, 4\%, and 1\%, respectively. These uncertainties were treated as independent of one another and combined in quadrature.

\subsubsection{Charged/neutral and baryon/meson ratios in MC training dataset} \label{s:chargedneutral}

In addition to the input $dN/d\eta$, we found that the charged/neutral and baryon/meson ratios in the MC dataset also slightly influence the unfolded results. To estimate the systematic uncertainty, these ratios were separately modified as functions of $\eta$ to match the shape at a higher or lower collision energy, similarly to how $dN/d\eta$ was re-shaped. At $\sqrt{s_{_{\text{NN}}}} =$ 200 and 7.7~GeV, only one adjacent energy is available, so instead, we used the uncertainties obtained at $\sqrt{s_{_{\text{NN}}}} =$ 27 and 9.2~GeV, respectively. The average uncertainty derived from varying the charged/neutral and baryon/meson ratios was 2\% and 3\%, respectively.

\subsubsection{Charged fraction correction method} \label{s:chargedcorrmethod}

The charged fraction correction can be performed using three different methods. The default method was described in Sec.~\ref{s:unfolding}.

In the second method, the charged fraction ($N_{\text{ch}}(i_{\text{ring}})/N_{\text{all}}(i_{\text{ring}})$), i.e., the fraction of EPD hits originating from charged primaries, is calculated for each EPD ring. The measured  hit numbers, $N(i_{\text{Ring}})$, are then multiplied by $N{_\text{ch}}(i_{\text{ring}})/N_{\text{all}}(i_{\text{ring}})$, and the unfolding is performed using a response matrix constructed only from charged particles. After unfolding, a multiple counting correction is applied, with the correction factor also calculated using only charged particles.

The third method utilizes \texttt{RooUnfold}'s built-in \texttt{Fake} method. Here, EPD hits originating from neutral primaries are treated as ``fake'' hits and are filled into the response matrix using the \texttt{Fake} method, which automatically accounts for the charged fraction correction. After unfolding, a multiple counting correction is applied, but the correction factor is again calculated using only charged particles.

The average uncertainty obtained using the different methods was around 1\%.

\subsubsection{Landau fit method} \label{s:fitmethod}

To estimate the systematic uncertainty arising from the choice of fitting method, fits were performed using two alternative approaches. In the first one, the Gaussian and Landau distributions were convoluted over the interval $(-\infty,\infty)$, i.e., a two-sided Gaussian was used to account for the broadening of the spectra. In the second alternative approach, the traditional fitting method without the Poisson constraint on the multi-MIP weights was employed. This method fails when an alpha peak is present, so in such cases the uncertainty was estimated as the average of the rings that do not exhibit an alpha peak. In both alternative cases, the resulting EPD hit numbers were unfolded using the default response matrices. On average, the systematic uncertainty from this estimate of the Landau fit error was around 7\%.

When alphas are present, it is possible for singly-charged particles and alphas to hit the same ring. This is mainly important for the innermost ring where the tiles are larger and the number of alpha particles is significant. To account for this, an extra alternative fitting method was applied in cases where the alpha weight in the default fit was nonzero. In this method, the mean number of alpha particles and the mean number of singly-charged particles were used as fit parameters. In addition to the n-MIP peaks and the alpha peak, two extra peaks were included, corresponding to cases where one alpha particle and either one or two singly-charged particles hit the same tile. The weight for each peak was calculated as the product of two Poisson distributions, with the mean number of alphas and singly-charged particles serving as the Poisson parameters. This alternative fitting method consistently resulted in a larger singly-charged yield. The difference compared to the default method reached about 10\% in the innermost ring at the lowest energies, but decreased rapidly toward the outer rings and the larger energies. The obtained uncertainty was added in quadrature to that originating from the choice of fitting method discussed in the previous paragraph.

\subsubsection{Number of Bayesian iterations} \label{s:numofit}

Naively, one might think that a larger number of iterations during Bayesian iterative unfolding leads to more accurate results. While this is true to some extent, too many iterations can amplify random fluctuations. The recommended number of iterations is 4~\cite{unfold_theor_new}, but we found that the unfolding converges reasonably well after 3 iterations. In some cases, random fluctuations become more pronounced in the 4th iteration. Ultimately, we decided that the default number of iterations would be 3, with $\pm1$ iteration used to estimate the systematic uncertainty, that turned out to be around 1\% on average.

\subsubsection{Impact parameter for MC}

The desired centrality for a MC simulation is achieved by specifying the impact parameter of the collision. The average impact parameter values for each centrality class can be calculated using the Glauber model. In this analysis, the impact parameter values from Ref.~\cite{phenixglauber}, calculated for $\sqrt{s_{_{\text{NN}}}} = 200$~GeV Au+Au collisions, were used. To account for potential systematic uncertainty arising from the choice of impact parameters, we performed the unfoldings using response matrices from adjacent centrality classes. For example, for the 10-20\% centrality class, we unfolded using response matrices derived from 5-10\% and 20-30\% centrality MC events. The average systematic uncertainty obtained this way was around 3\%.

\subsubsection{East-West asymmetry}


We observe a small violation of the symmetry $dN(\eta)/d\eta = dN(-\eta)/d\eta$ in our extracted spectra. This asymmetry is smaller in central collisions (around 2\%) and increases for more peripheral cases, reaching around 10\% in the most peripheral class. The systematic uncertainty resulting from the East-West asymmetry was quantified as the difference between the unfolded $dN_{\text{ch}}/d\eta$ at $\eta$ and $-\eta$. At $\sqrt{s_{_{\text{NN}}}} =$ 54.4~GeV, the uncertainty obtained at $\sqrt{s_{_{\text{NN}}}} =$ 27~GeV was utilized.

\subsubsection{GEANT hadronic interaction package}

There are two hadronic interaction packages available in GEANT: GHEISHA and FLUKA~\cite{geant}. 
The choice of package has a small effect on the response matrix, and therefore on the final results. 
For this analysis, GHEISHA was used as the default option, while FLUKA was employed to estimate the systematic uncertainty. The difference turned out to be on the order of 3\%.

\subsubsection{Alpha-induced secondaries}

The calculation of the number of alpha-induced secondaries relies on our evaluation of the primary alpha-particle distribution. The dominant uncertainty on the extracted alpha-particle yield arises from the Landau fits. This varies point-to-point, with an average of about 13\%. To include other smaller effects (e.g., extrapolation outside the measured range, etc.) we increase this to a conservative value of 20\%, and for the systematic, varied the distribution by this amount. The resulting systematic uncertainty on the final $dN_{\text{ch}}/d\eta$ reached up to about 7\% in cases where the correction for alpha-induced secondaries was significant. However, similarly to the correction itself, it decreased rapidly with increasing collision energy and outside the spectator $\eta$ region.

\subsubsection{Summed systematics}

An asymmetric systematic uncertainty was calculated for each source of uncertainty outlined above. Unless otherwise noted, the uncertainties were computed on a bin-by-bin basis, separately for each collision energy and centrality class. The uncertainty sources were treated as independent, and the uncertainty values were combined in quadrature. Overall scale uncertainties and point-by-point uncertainties were not separated. The upper and lower uncertainties from each source, averaged over centrality classes and $\eta$ bins, are presented in Table~\ref{t:syst} for each collision energy.

\subsubsection{Systematics summary}

The total systematic uncertainty is on the order of 10\%. 
This value is roughly independent of collision energy but depends on centrality and $\eta$. 
More peripheral centrality classes tend to have larger uncertainties, and the uncertainty also increases toward the edges of the EPD acceptance. 
The total uncertainty is generally not dominated by a single source but is instead a combination of several sources. 
Statistical uncertainties are always smaller and, in most cases, negligible compared to the systematic uncertainties.

\end{multicols}

\begin{table}[htbp]
\centering
\scriptsize
\setlength{\tabcolsep}{4pt}
\renewcommand{\arraystretch}{1.1}
\begin{tabular}{l*{8}{R{1.5cm}}}
\toprule
\textbf{Test} & \textbf{7.7 GeV} & \textbf{9.2 GeV} & \textbf{11.5 GeV} & \textbf{14.6 GeV} & \textbf{19.6 GeV} & \textbf{27 GeV} & \textbf{54.4 GeV} & \textbf{200 GeV} \\
\midrule
Multiplicity shift & 3.65/1.32 & 2.92/1.80 & 3.08/2.22 & 3.34/2.15 & 3.28/2.62 & 3.72/2.66 & 4.13/2.92 & 5.55/3.78 \\
Vz resolution & 0.74/0.74 & 0.74/0.74 & 0.74/0.74 & 0.74/0.74 & 0.74/0.74 & 0.74/0.74 & 0.73/0.74 & 0.70/0.72 \\
Vz choice & 2.22/3.02 & 2.22/3.02 & 2.22/3.02 & 2.22/3.02 & 2.22/3.02 & 2.22/3.02 & 2.24/3.01 & 1.71/2.82 \\
$dN/d\eta$ wide/narrow & 3.49/6.88 & 3.46/6.76 & 3.70/6.77 & 4.18/6.91 & 4.46/6.81 & 4.50/6.33 & 4.11/5.39 & 3.14/4.54 \\
$dN/d\eta$ scaling & 3.01/3.27 & 3.01/3.27 & 3.44/3.35 & 3.25/2.77 & 3.24/2.68 & 5.74/3.36 & 6.15/4.46 & 5.61/2.14 \\
Scale $dN/d\eta$ to unfolded & 2.08/0.87 & 1.49/0.76 & 1.08/0.63 & 0.80/0.58 & 0.70/0.56 & 0.79/0.78 & 0.47/0.64 & 1.82/1.05 \\
charged/neutral & 1.59/0.93 & 1.59/0.93 & 1.59/1.33 & 1.86/1.31 & 1.99/1.60 & 3.70/1.67 & 3.83/3.56 & 3.58/1.80 \\
baryon/meson & 1.26/0.84 & 1.26/0.84 & 1.36/0.95 & 1.53/1.03 & 1.81/1.21 & 3.96/1.35 & 5.20/4.35 & 4.19/1.51 \\
Charged fraction corr. & 1.11/0.47 & 0.83/0.55 & 0.56/0.65 & 0.57/0.91 & 0.76/0.95 & 0.79/0.78 & 0.47/0.64 & 1.82/1.05 \\
Landau fit & 8.30/0.54 & 8.42/1.51 & 7.37/1.63 & 7.77/2.41 & 7.07/2.43 & 6.31/2.16 & 5.15/3.33 & 5.99/2.10 \\
\#Bayesian iterations & 1.23/1.32 & 0.93/1.01 & 0.69/0.74 & 0.73/0.75 & 0.65/0.65 & 0.60/0.57 & 0.56/0.51 & 0.51/0.67 \\
MC impact par. & 3.62/2.96 & 3.26/2.55 & 3.10/2.34 & 3.42/2.63 & 3.65/2.91 & 3.35/2.75 & 2.55/2.32 & 0.93/0.91 \\
East-West asym. & 2.13/2.00 & 2.33/2.18 & 2.68/2.48 & 3.20/2.92 & 3.42/3.11 & 3.17/2.94 & 6.30/0.05 & 4.34/3.89 \\
Hadronic package & 5.96/0 & 5.27/0.01 & 3.76/0.03 & 3.03/0.18 & 2.31/0.57 & 2.25/0.74 & 1.61/0.52 & 1.31/0.32 \\
Alpha secondaries & 0.60/1.67 & 0.58/1.60 & 0.58/1.07 & 0.64/0.68 & 0.76/0.40 & 0/0 & 0/0 & 0/0 \\
Total & 13.18/9.47 & 12.57/9.43 & 11.53/9.55 & 12.00/9.80 & 11.67/10.01 & 13.01/9.76 & 13.86/10.77 & 12.92/8.71
 \\
\bottomrule
\end{tabular}
\caption{Systematic uncertainties (\%) averaged over $\eta$ bins and centrality classes for various systematics considerations across different $\sqrt{s_{_{\text{NN}}}}$. Each cell shows the upper and lower uncertainties from each source.}
\label{t:syst}
\end{table}

\newpage
\begin{multicols}{2}

\section{Results and discussion} \label{s:results}

\begin{figure}[p]
    \centering
    \includegraphics[width=0.49\linewidth,trim={0 0 0 1cm},clip]{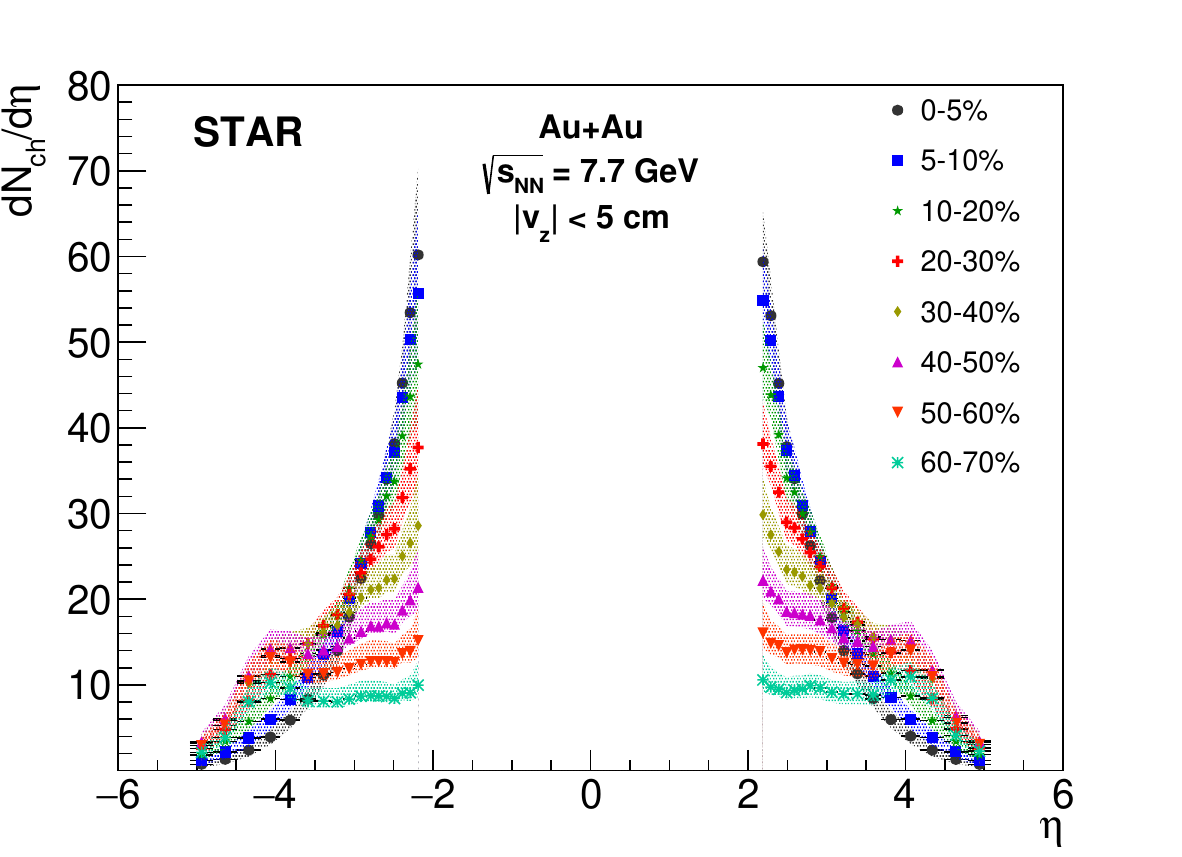}
    \includegraphics[width=0.49\linewidth,trim={0 0 0 1cm},clip]{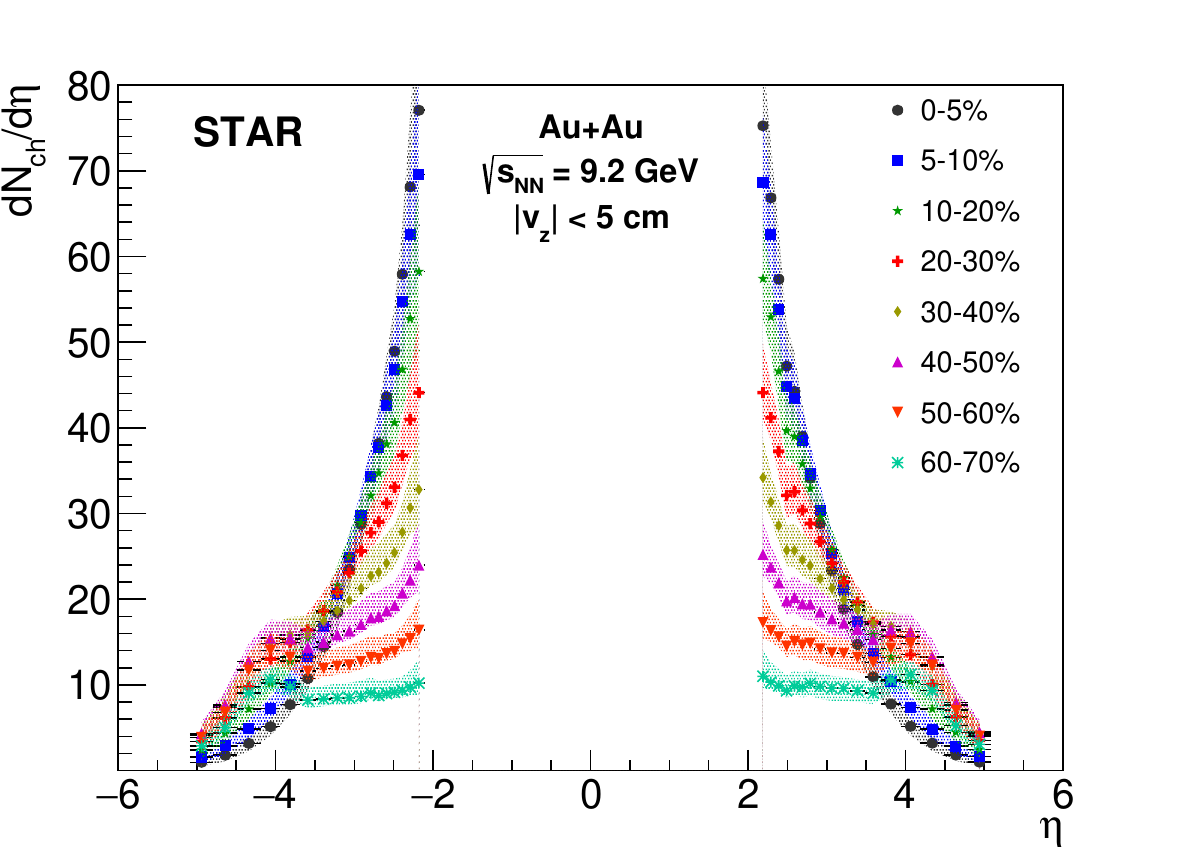}
    \includegraphics[width=0.49\linewidth,trim={0 0 0 1cm},clip]{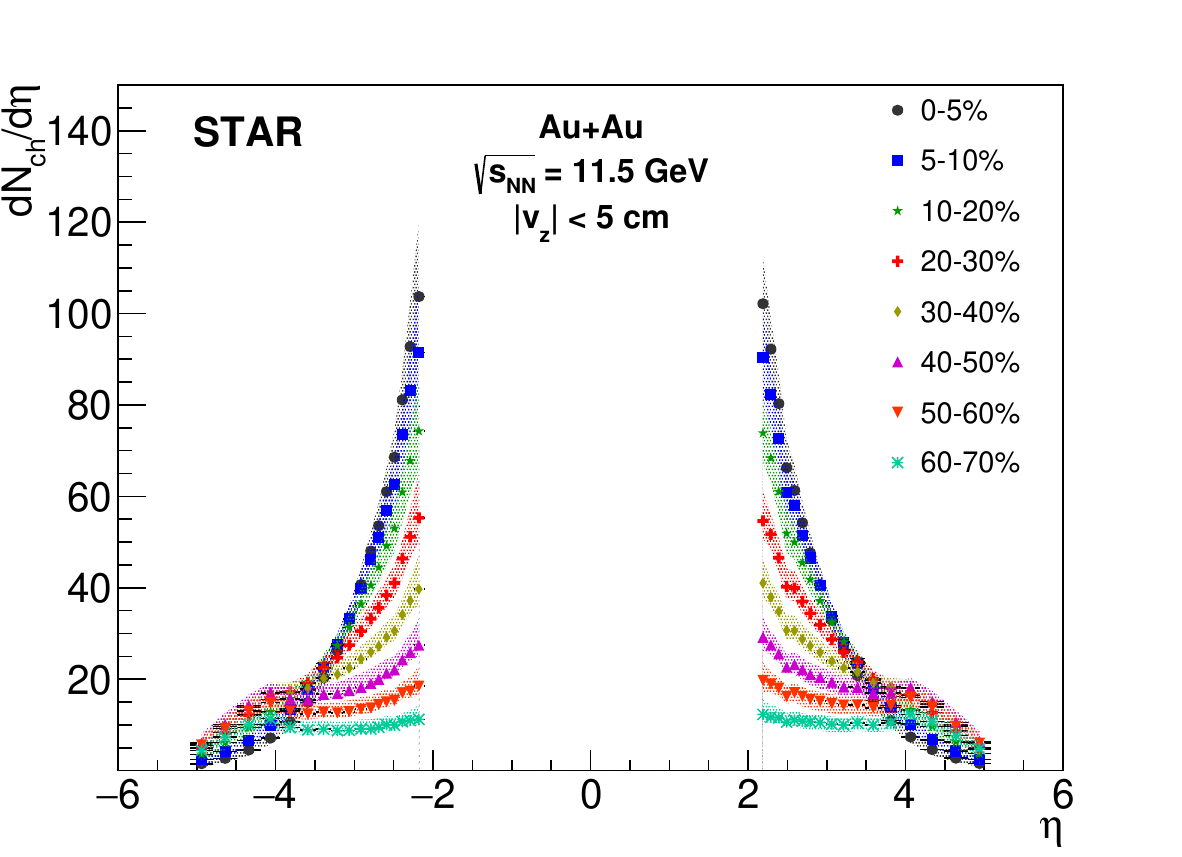}
    \includegraphics[width=0.49\linewidth,trim={0 0 0 1cm},clip]{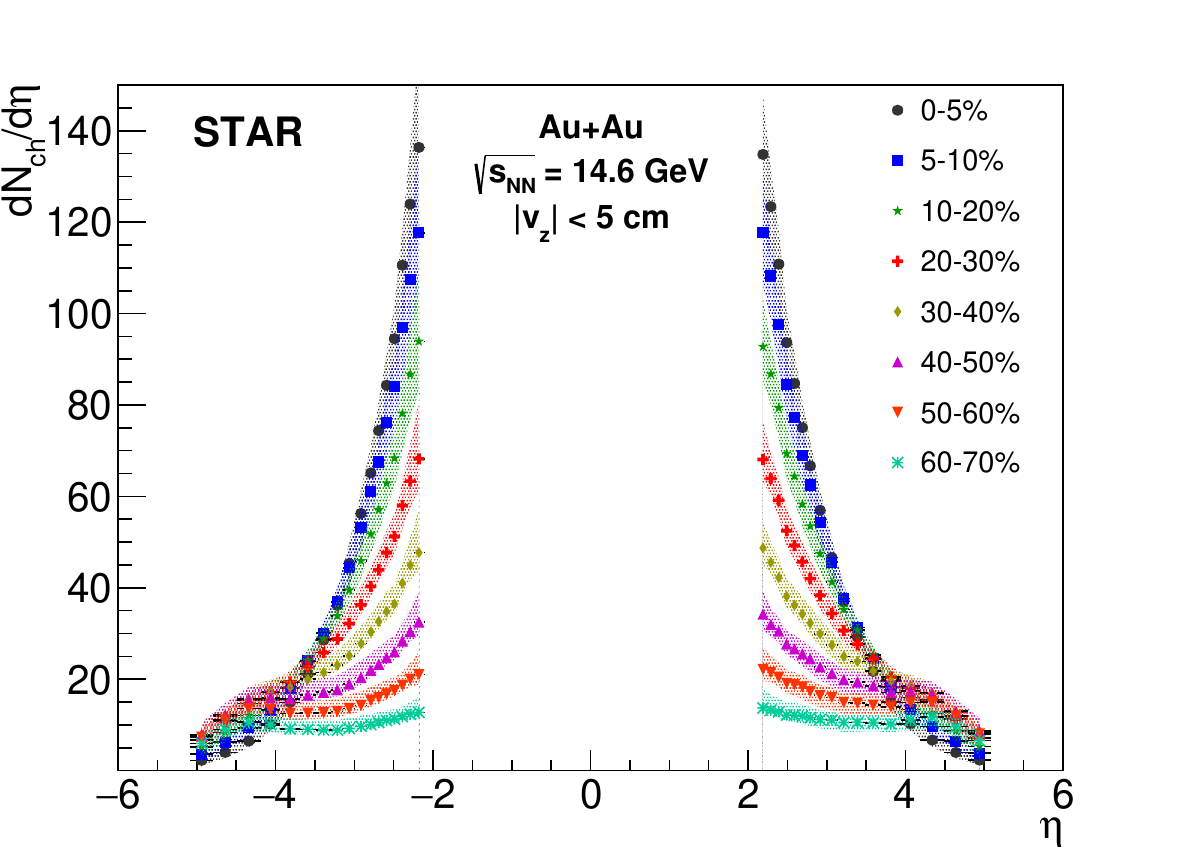}
    \includegraphics[width=0.49\linewidth,trim={0 0 0 1cm},clip]{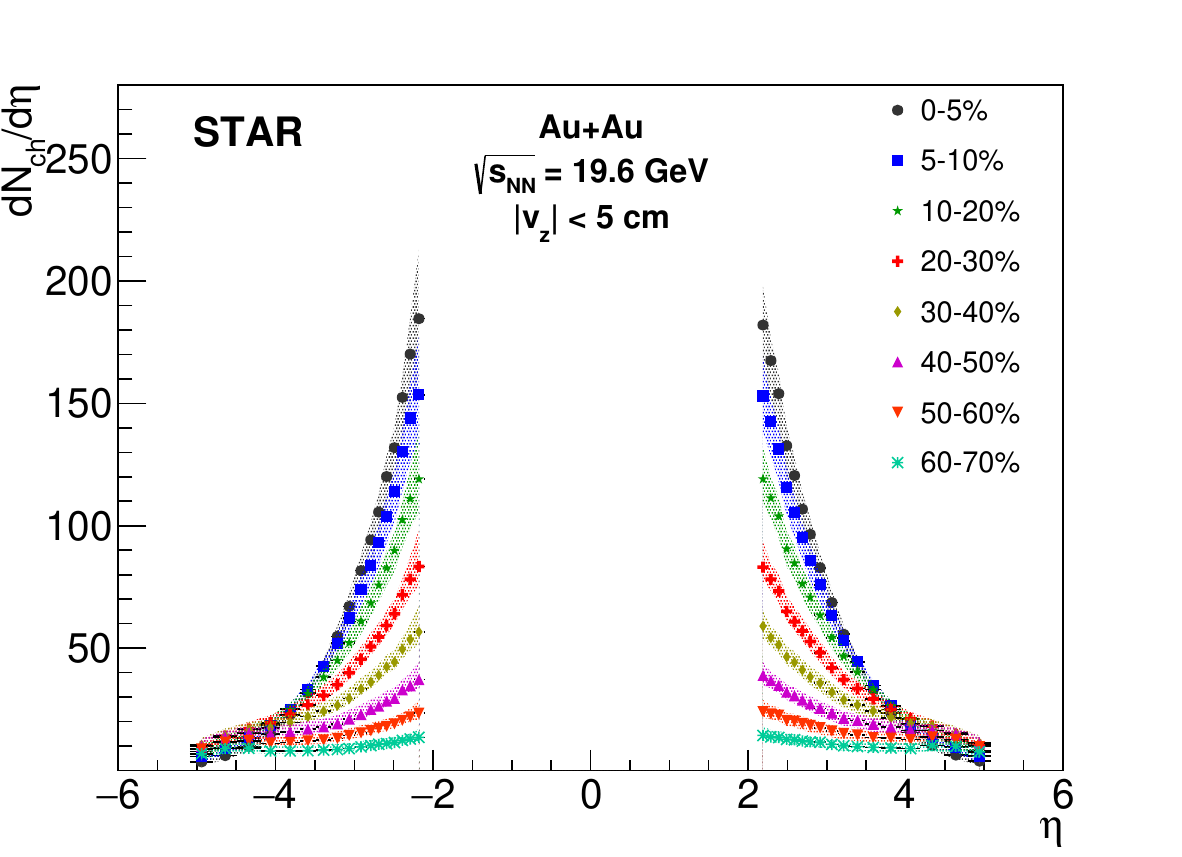}
    \includegraphics[width=0.49\linewidth,trim={0 0 0 1cm},clip]{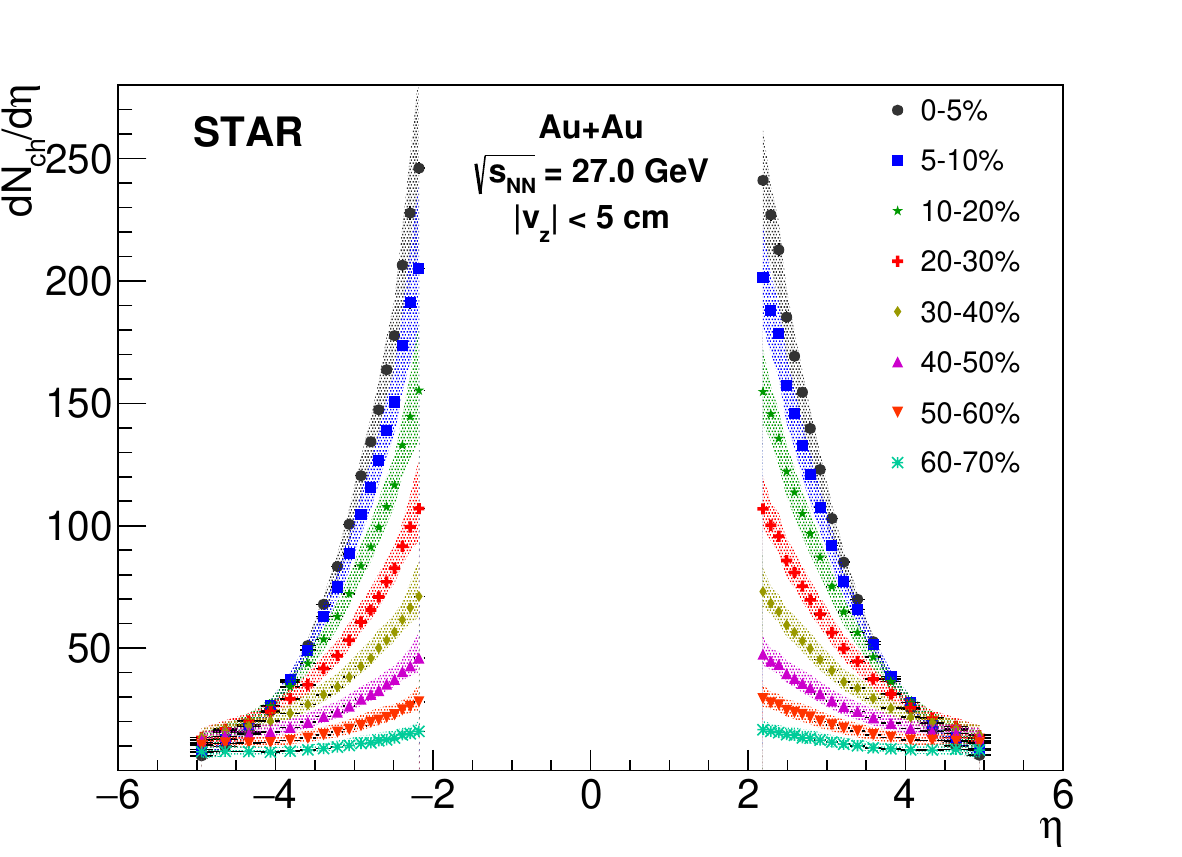}
    \includegraphics[width=0.49\linewidth,trim={0 0 0 1cm},clip]{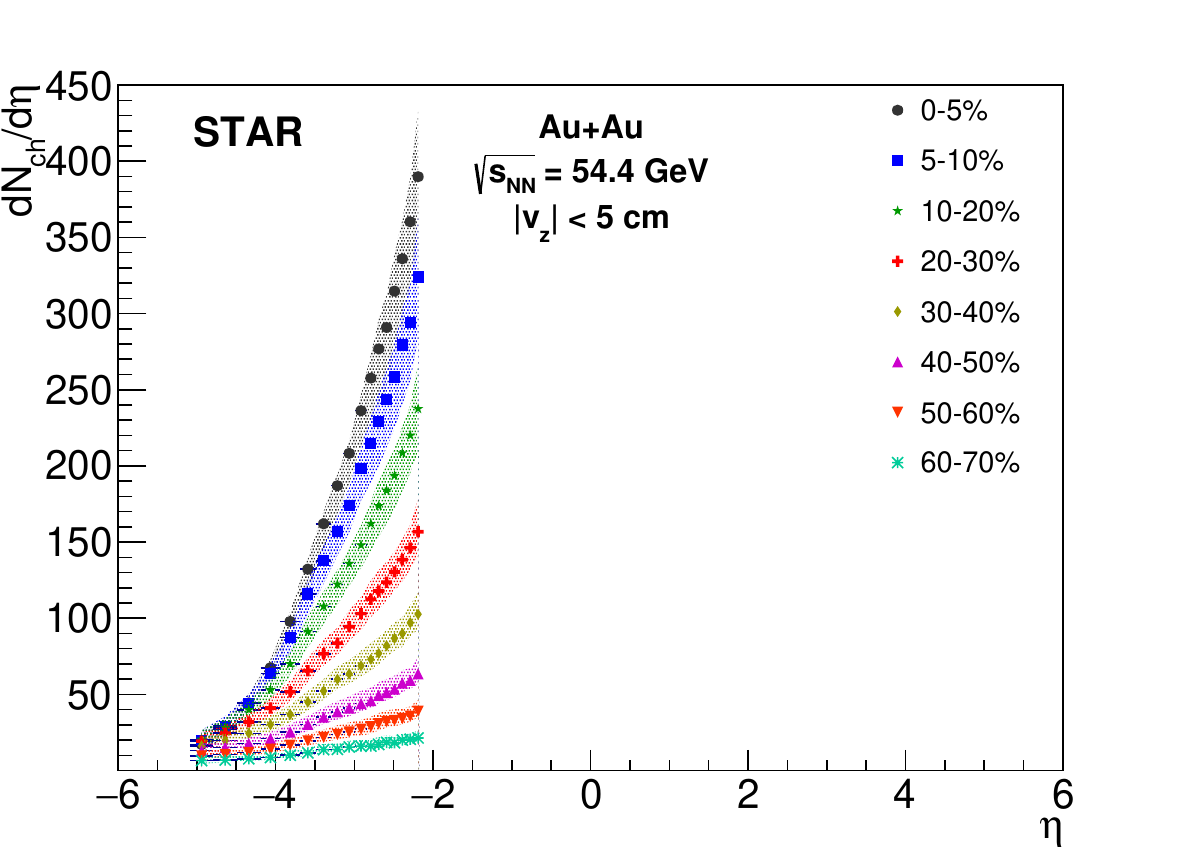}
    \includegraphics[width=0.49\linewidth,trim={0 0 0 1cm},clip]{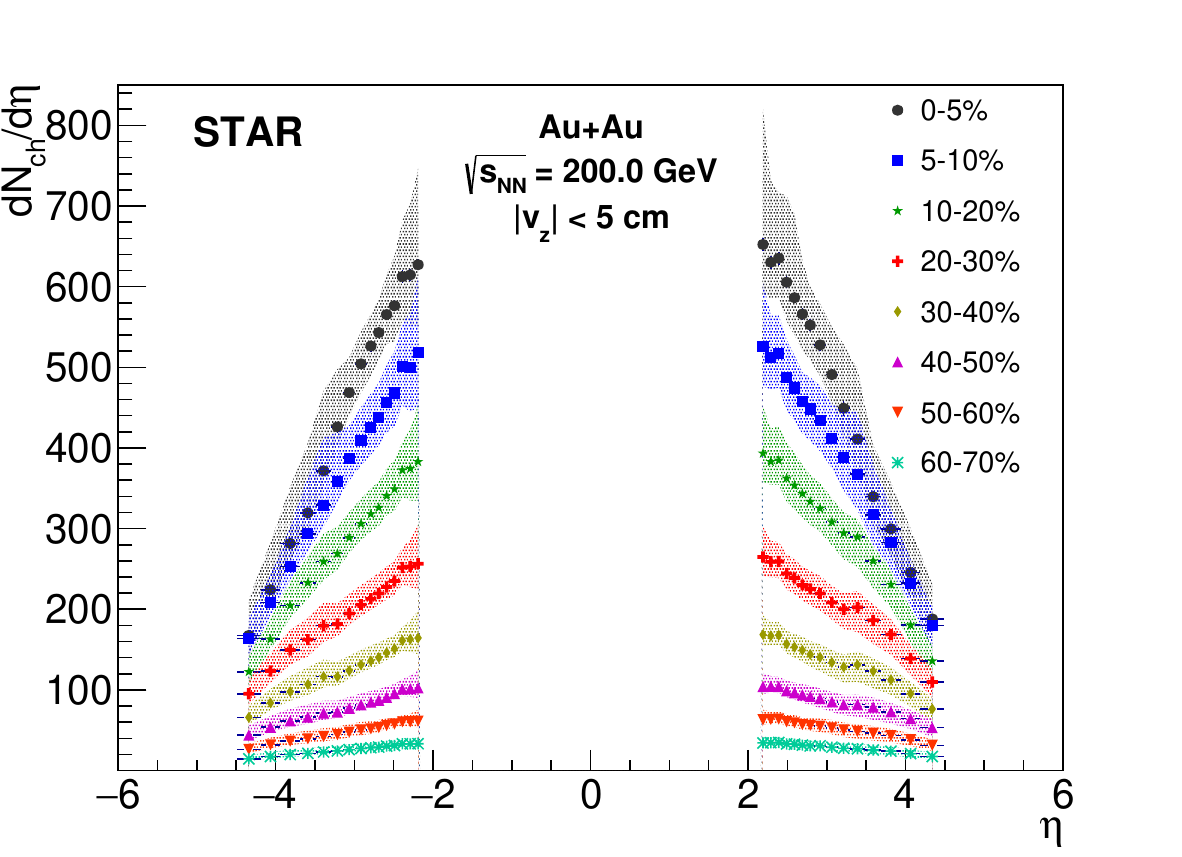}
    \caption{Pseudorapidity distribution of charged particles ($dN_{\text{ch}}/d\eta$) measured by the EPD at (from left to right, top to bottom) $\sqrt{s_{_{\text{NN}}}}=$ 7.7, 9.2, 11.5, 14.6, 19.6, 27, 54.4, and 200~GeV for different centrality classes. Shaded bands correspond to systematic uncertainties. Statistical uncertainties are shown as error bars, but they are usually smaller than the markers. At $\sqrt{s_{_{\text{NN}}}}=$ 54.4~GeV, due to partial installation, results are only measured for the $\eta<0$ side.}
    \label{f:final}
\end{figure}

The measured charged particle pseudorapidity distributions are shown in Fig.~\ref{f:final} for $\sqrt{s_{_{\text{NN}}}} =$ 7.7, 9.2, 11.5, 14.6, 19.6, 27, 54.4, and 200~GeV.  As expected, the general trends indicate that $dN_{\text{ch}}/d\eta$ increases with collision energy, toward midrapidity, and in more central collisions. At $\sqrt{s_{_{\text{NN}}}} = 27$~GeV and lower energies, a ``shoulder'' appears at large $|\eta|$, with its relative size increasing in more peripheral collisions and at smaller energies. The shoulder also moves towards larger $|\eta|$ as the collision energy increases. These shoulders are attributed to protons coming from spectators, meaning that only models including spectators can reasonably reproduce the experimental data at these energies. Simple hydrodynamic models, which apply only to participants, are therefore only applicable to the most central collisions, where spectator contributions are negligible. 

\begin{figure}[h!]
    \centering
    \includegraphics[width=0.49\linewidth]{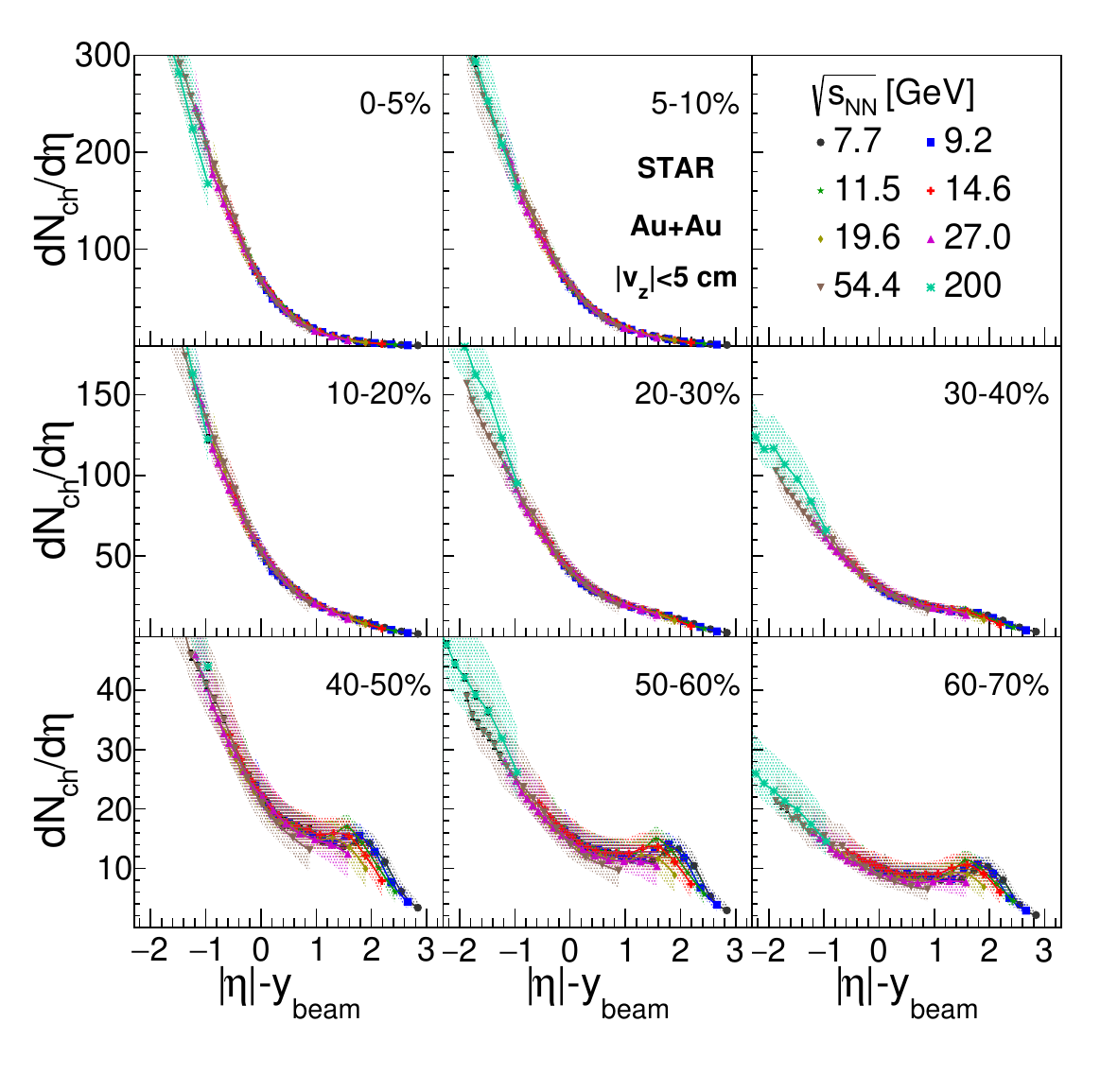}
    \includegraphics[width=0.49\linewidth]{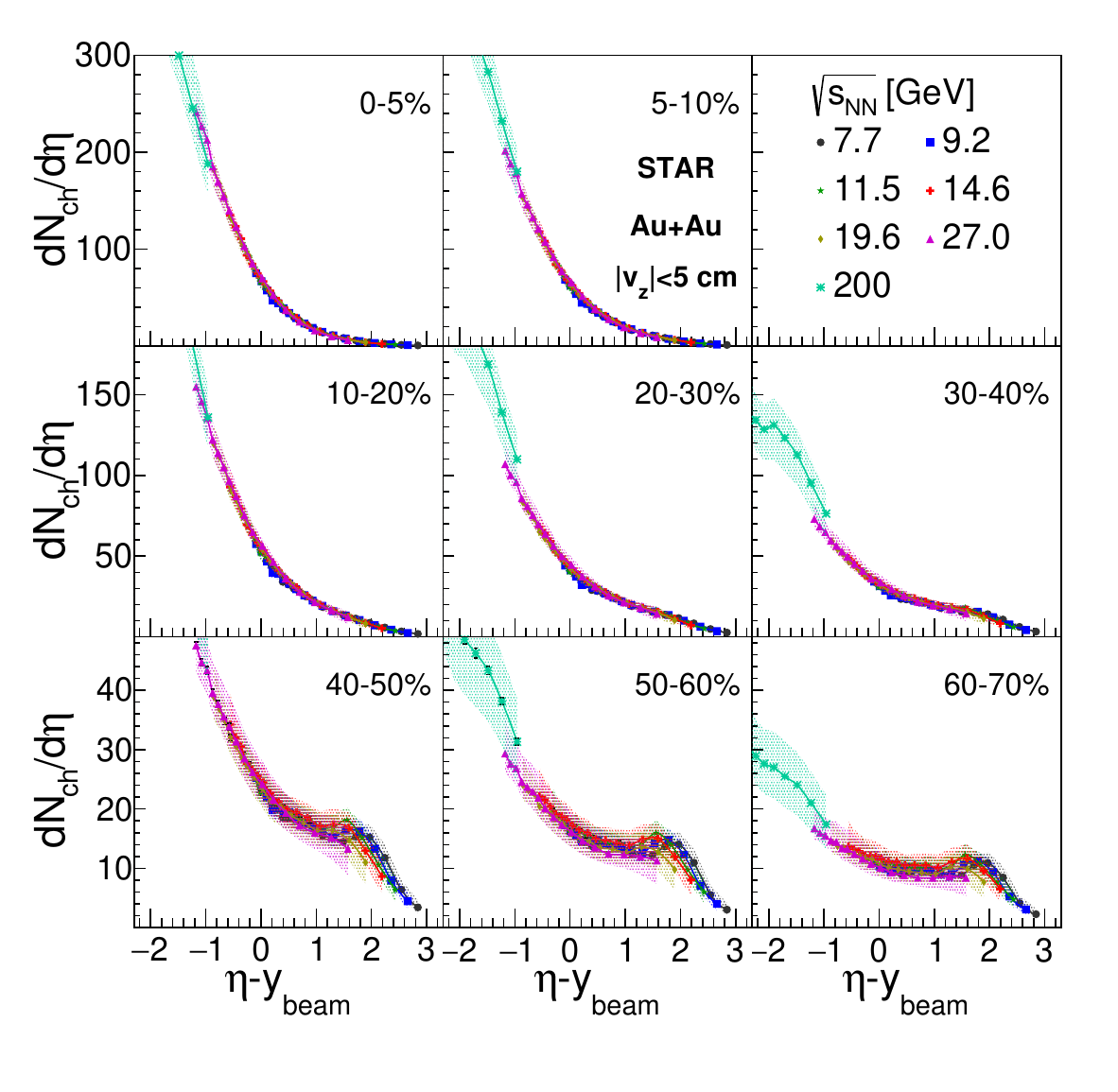}
    \caption{Pseudorapidity distribution of charged particles ($dN_{\text{ch}}/d\eta$) shifted by the beam rapidity ($y_{\text{beam}}$) for the negative (left) and positive (right) $\eta$ regions. For relativistic particles $\eta\approx y$, thus $\eta-y_{\text{beam}}$ is a proxy for the pseudorapidity in the rest frame of one of the colliding nuclei. Shaded bands correspond to systematic uncertainties. Statistical uncertainties are smaller than the markers.}
    \label{f:limitingfragmentation}
\end{figure}

At sufficiently high collision energies, it is well established that the particle production in the rapidity region associated with either beam becomes nearly independent of the collision energy~\cite{Benecke:1969sh}. This behavior, known as limiting fragmentation scaling, has been observed in $p+p$, $p+A$, and heavy-ion collisions~\cite{PHOBOS:2010eyu}. The scaling extends over a wider range in pseudorapidity than expected from the simple limiting-fragmentation picture and appears to apply to other observables in heavy-ion collisions, such as directed flow~\cite{ STAR:2024ujm}. Nonetheless, the underlying physical mechanisms behind limiting fragmentation still remain unclear. Although the original formulation of this concept refers to particle production as a function of rapidity and transverse momentum, it can be shown that a similar scaling behavior is also present in $dN_{\text{ch}}/d\eta$~\cite{PHOBOS:2010eyu}. Figure~\ref{f:limitingfragmentation} shows $dN_{\text{ch}}/d\eta$ shifted by the beam rapidity,
\begin{equation}
    y_{\text{beam}}=\cosh^{-1}\left(\frac{\sqrt{s_{_{\text{NN}}}}}{2m_N} \right),
\end{equation}
where $m_N$ is the nucleon mass. In central collisions, the results follow the same curve across different collision energies, supporting the validity of the limiting fragmentation hypothesis.  However, limiting fragmentation breaks down towards more peripheral collisions due to the presence of spectators. The shoulder associated with spectators shifts to smaller values of $\eta$ as the energy decreases; however, in terms of $|\eta| -y_{\text{beam}}$, it appears at larger values for lower energies. Although the systematic uncertainties overlap even in the spectator-dominated region, they are strongly correlated across different energies.

\begin{figure}[h!]
    \centering
    \includegraphics[width=0.49\linewidth]{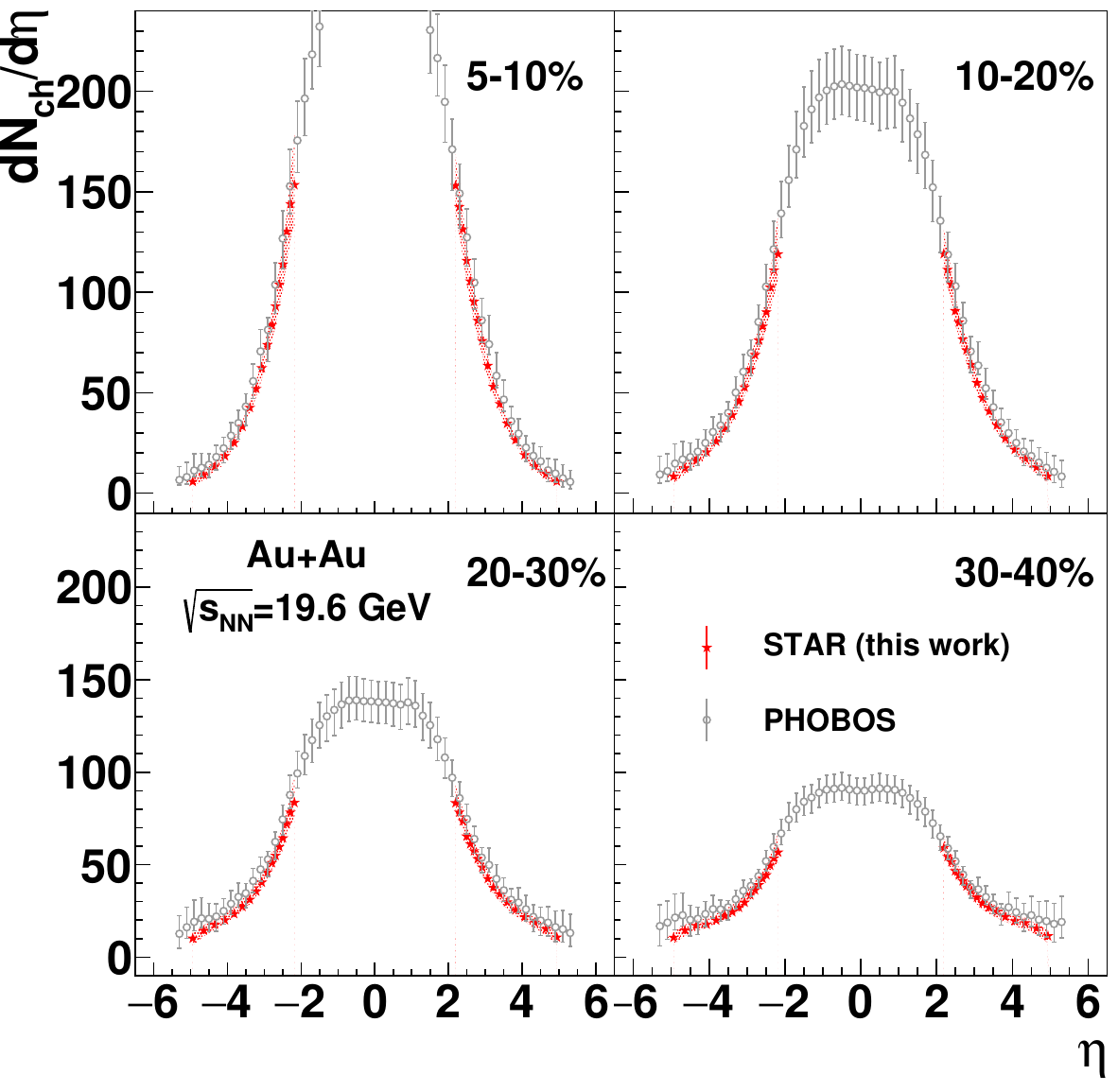}
    \includegraphics[width=0.49\linewidth]{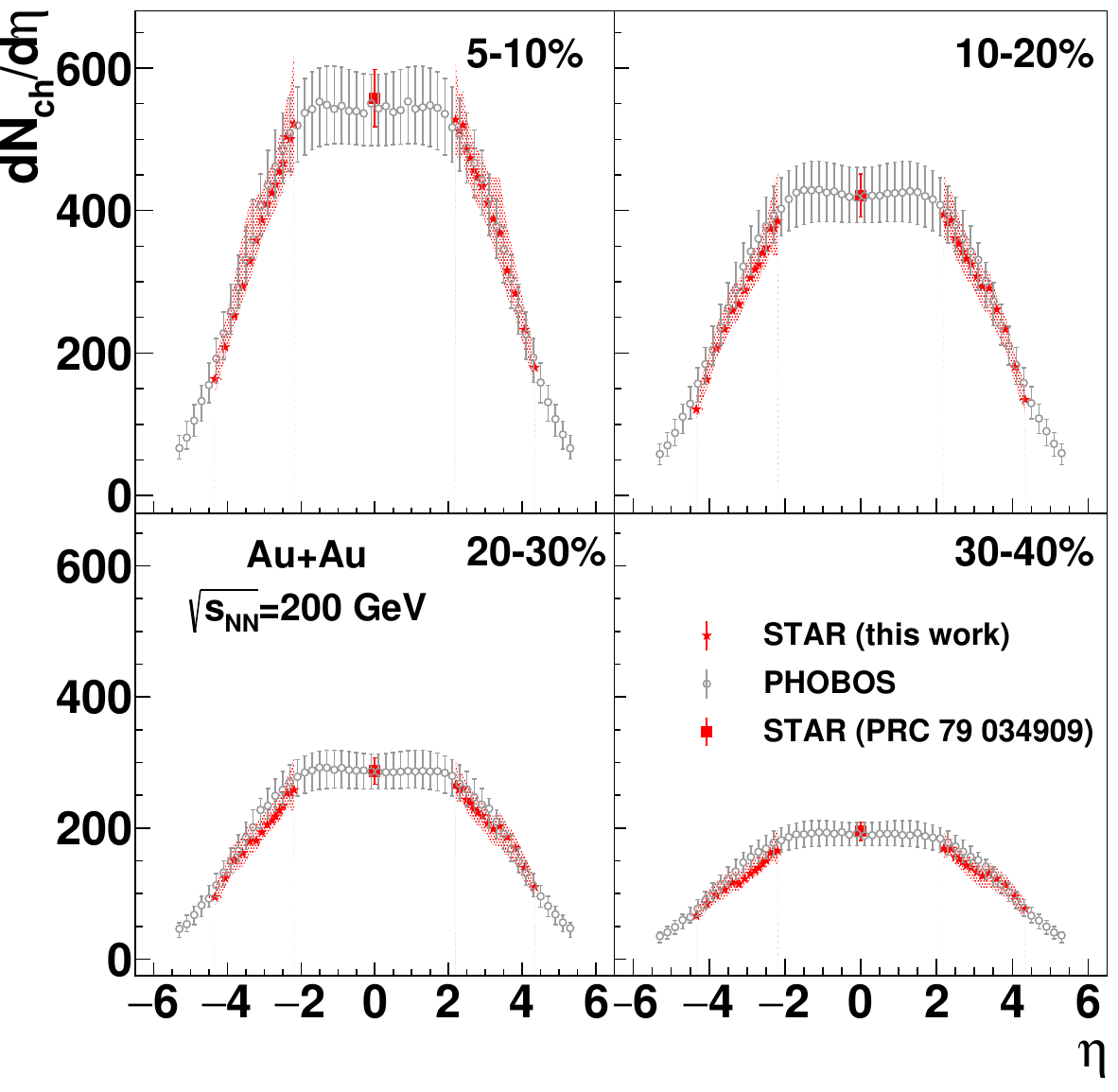}
    \caption{Comparison between the pseudorapidity distribution of charged particles ($dN_{\text{ch}}/d\eta$) measured by STAR and PHOBOS at $\sqrt{s_{_{\text{NN}}}} = 19.6$~GeV (left) and $200$~GeV (right) across four centrality classes. PHOBOS and STAR results are shown with open and solid markers, respectively. The error bars on the PHOBOS data and the previously published STAR results represent the combined statistical and systematic uncertainties.  Systematic uncertainties of the new STAR data are indicated by shaded bands; statistical uncertainties are smaller than the markers.}
    \label{f:phobos}
\end{figure}

The PHOBOS Collaboration measured $dN_{\text{ch}}/d\eta$ in Au+Au collisions at various energies and centralities~\cite{PHOBOS:2010eyu}. Two common energies compared to this analysis are $\sqrt{s_{_{\text{NN}}}} = $ 19.6 and 200~GeV. The comparison of the STAR and PHOBOS results for these two energies is shown in Fig.~\ref{f:phobos}, across four common centrality classes. Note that PHOBOS used a 6–10\% centrality class rather than 5–10\%, and 5\%-wide classes between 10\% and 40\% that were averaged to match the 10\%-wide classes in this analysis.
The STAR results are in agreement with the PHOBOS data within uncertainties, although, the former are systematically smaller in some cases. The previously measured STAR data~\cite{STAR:2008med} at $\sqrt{s_{_{\text{NN}}}} = $ 200~GeV is in perfect agreement with PHOBOS at midrapidity. At $\sqrt{s_{_{\text{NN}}}} = $ 200~GeV the BRAHMS Collaboration also measured $dN_{\text{ch}}/d\eta$ over a wide range in $\eta$~\cite{BRAHMS:2001llo}, and its results are in agreement with PHOBOS.

\begin{figure}[h!]
    \centering
    \includegraphics[width=0.49\linewidth]{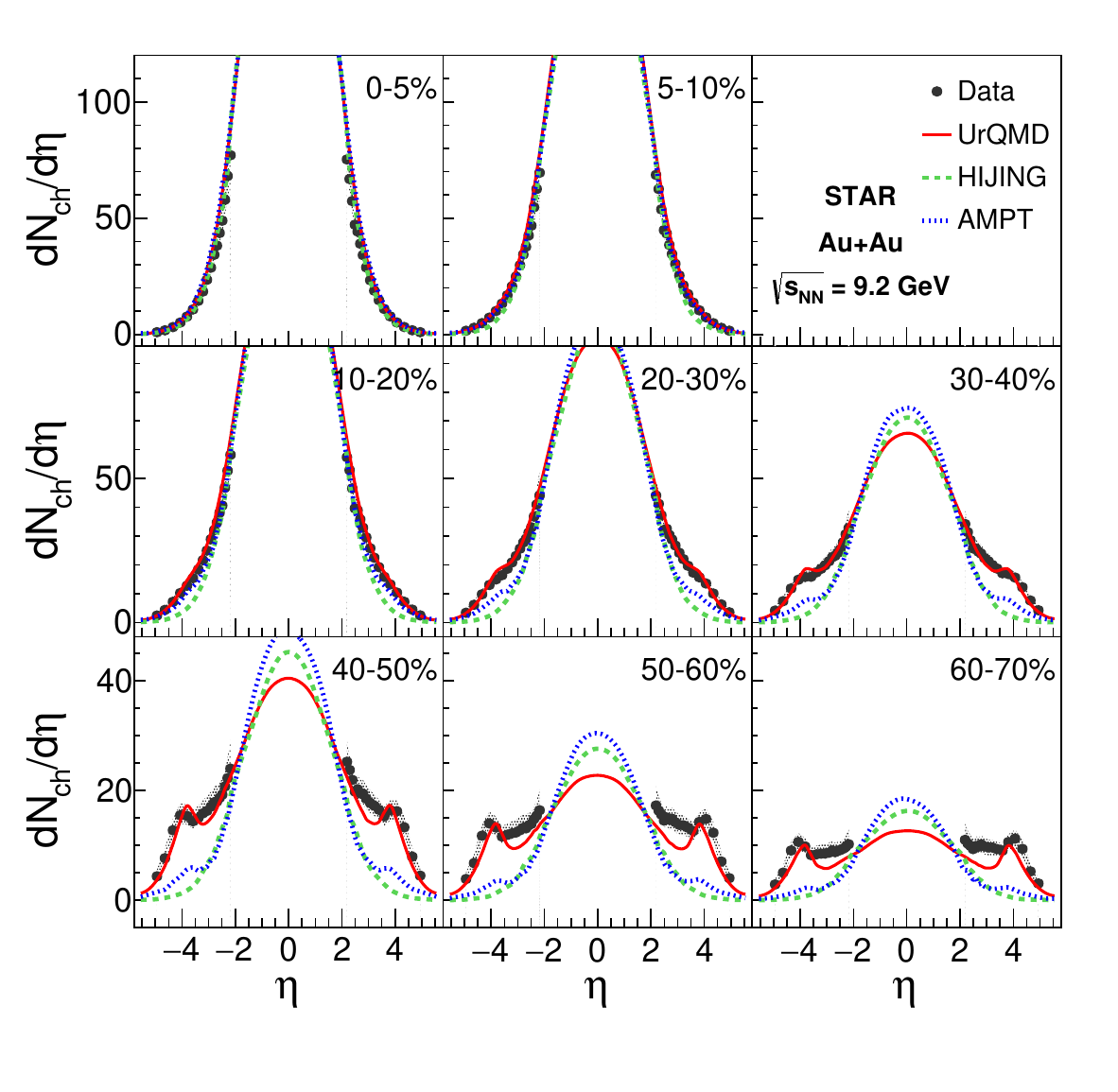}
    \includegraphics[width=0.49\linewidth]{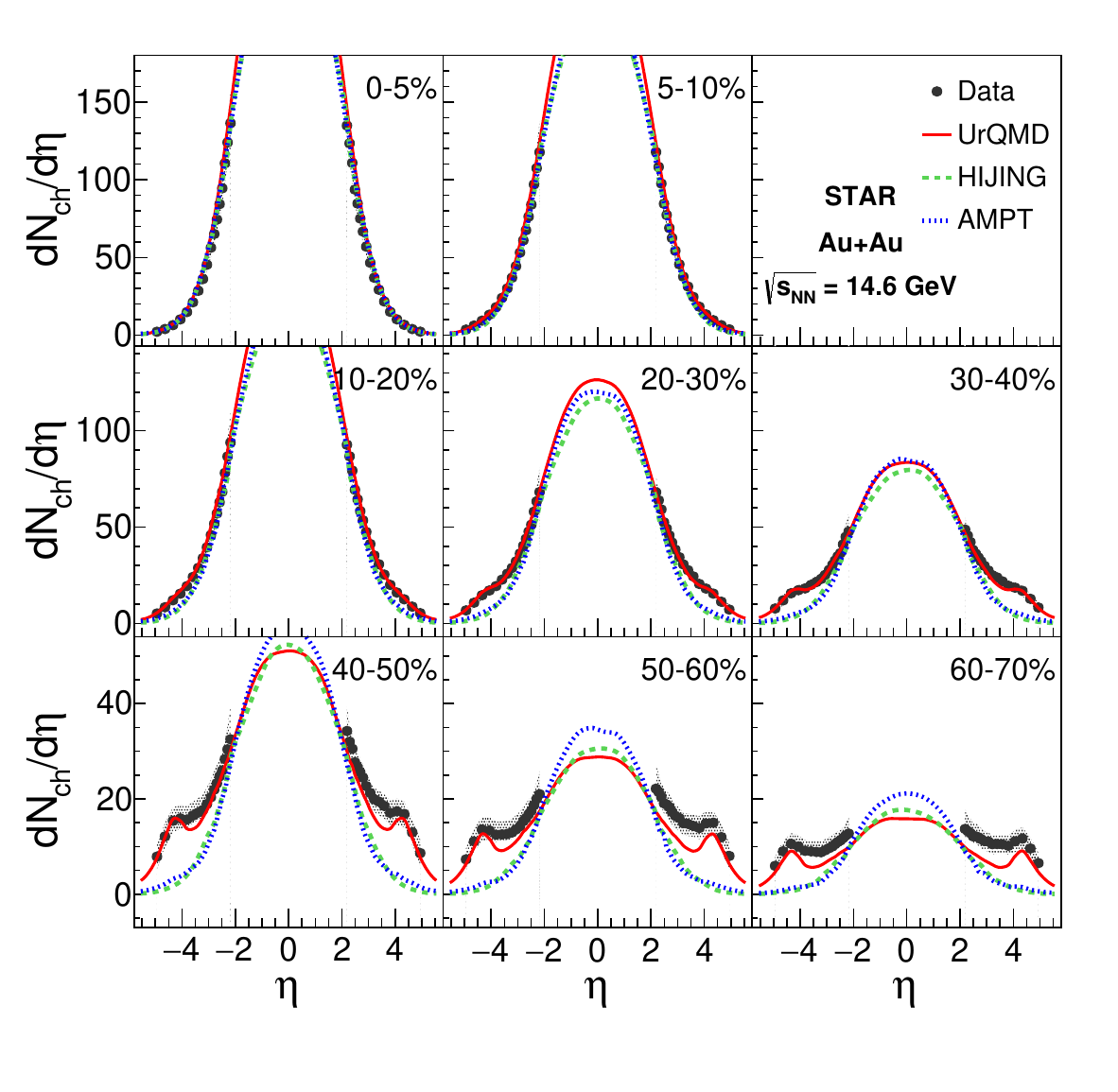}
    \includegraphics[width=0.49\linewidth]{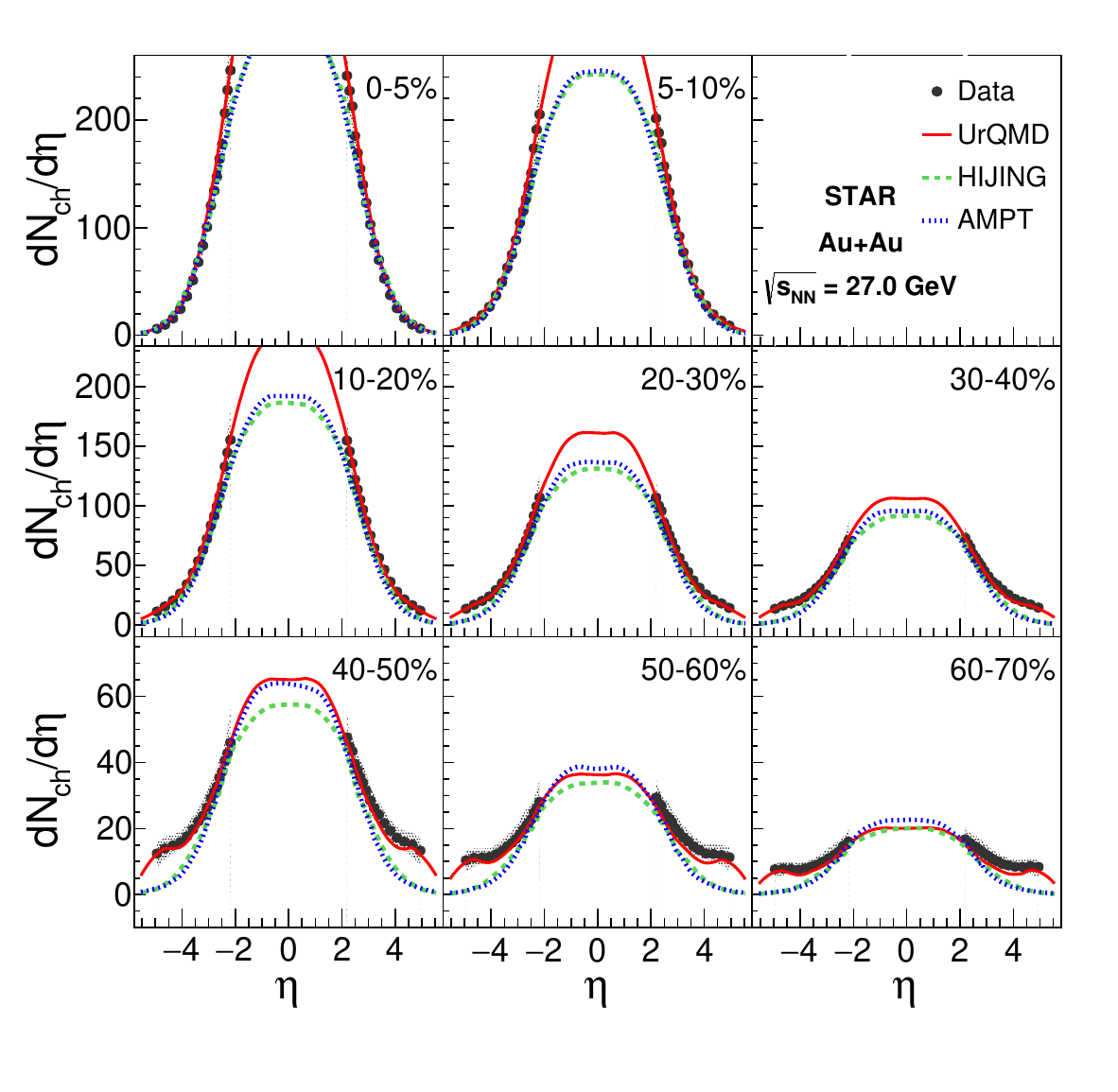}
    \includegraphics[width=0.49\linewidth]{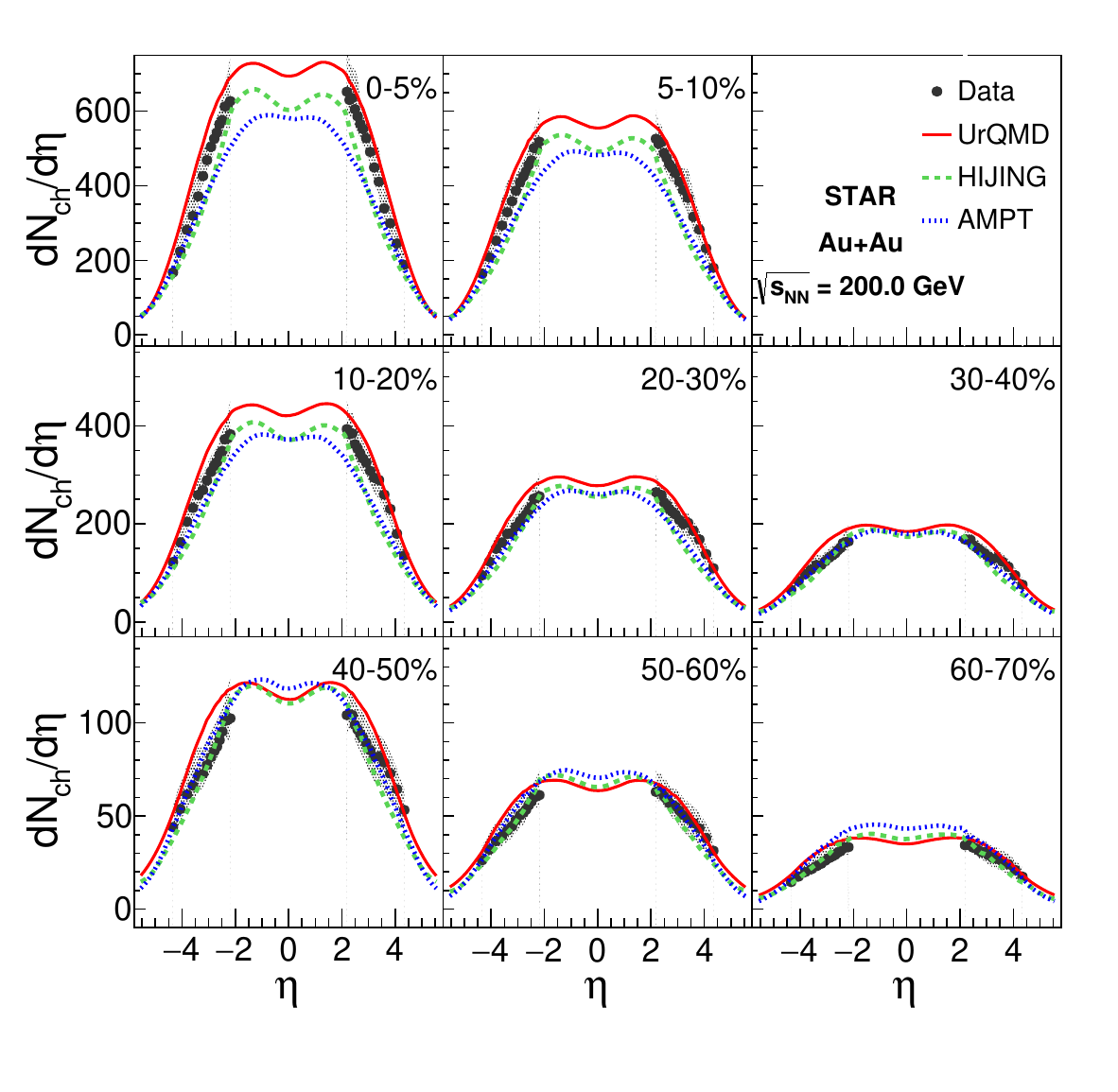}
    \caption{Comparison between pseudorapidity distributions of charged particles ($dN_{\text{ch}}/d\eta$) measured by STAR and predicted by models at $\sqrt{s_{_{\text{NN}}}}= $ 9.2 (upper left), 14.6 (upper right), 27 (lower left), and 200 (lower right)~GeV in different centrality classes. Model predictions are shown as lines. The STAR results are shown as full markers with (hardly visible) shaded bands corresponding to systematic uncertainties; statistical uncertainties are smaller than the markers.}
    \label{f:models}
\end{figure}

A comparison of the experimental results to three different models, UrQMD~\cite{urqmd1,urqmd2}, HIJING~\cite{hijing}, and AMPT~\cite{ampt} is shown in Fig.~\ref{f:models} for $\sqrt{s_{_{\text{NN}}}} = $ 9.2, 14.6, 27, and 200~GeV. UrQMD was used for the calculation of the response matrices, primarily because it includes spectators. At lower energies, both HIJING and AMPT deviate more from the data as one moves toward more peripheral collisions due to the absence of spectators in these models. In the most peripheral cases, even UrQMD fails to quantitatively describe the data. This may be because, although UrQMD includes spectators, it does not incorporate projectile fragmentation, i.e., it does not produce nuclear fragments with a mass number greater than one.  As the energy increases and the spectator contribution becomes less dominant, the differences between the models become less pronounced. However, at $\sqrt{s_{_{\text{NN}}}} = 27$~GeV, UrQMD still provides the best description. At $\sqrt{s_{_{\text{NN}}}} = 200$~GeV, UrQMD slightly overestimates the yield, while HIJING provides a better description of the data.

\end{multicols}

\newpage
\begin{multicols}{2}

\section{Summary} \label{s:summary}

The measurement of charged particle pseudorapidity distributions ($dN_{\text{ch}}/d\eta$) in gold-gold collisions at center-of-mass energies per nucleon pair of $\sqrt{s_{_{\text{NN}}}} = $ 200, 54.4, 27, 19.6, 14.6, 11.5, 9.2, and 7.7~GeV is presented using the STAR Event Plane Detector (EPD). The measurement is performed in 8 centrality classes ranging from 0\% to 70\% for the pseudorapidity range $2.14 < |\eta| < 5.09$ at each energy. For $\sqrt{s_{_{\text{NN}}}} = 54.4$~GeV, only the $\eta < 0$ side is measured.

The number of charged particles hitting each EPD ring is determined through convoluted Landau fits to the deposited energy spectra. The broadening of the MIP peaks in the spectra is accounted for by using a Landau form convoluted with a Gaussian distribution. During the fits, the Poisson distribution of n-MIP weights is enforced. Additionally, when present, alpha particles from spectators and their secondaries are taken into account.

The $dN_{\text{ch}}/d\eta$ distributions are extracted from the raw EPD hit numbers using Bayesian iterative unfolding. A response matrix is calculated for each collision energy and centrality class separately, using the UrQMD event generator and the GEANT3 simulation toolkit at the same energy and centrality as the corresponding data set. After unfolding, a multiple counting correction and a charged fraction correction are applied.

The unfolded results exhibit the expected general trends: $dN_{\text{ch}}/d\eta$ increases with collision energy, towards midrapidity, and in more central collisions. At energies below $\sqrt{s_{_{\text{NN}}}} = 27$~GeV, a contribution from spectators is visible in peripheral collisions at large $|\eta|$, with its relative size growing towards lower energies. Limiting fragmentation is fulfilled in central collision cases, but it breaks down as one moves toward more peripheral collisions due to the presence of spectators. Comparing the STAR results to $dN_{\text{ch}}/d\eta$ measured by the PHOBOS Collaboration at $\sqrt{s_{_{\text{NN}}}} = $ 19.6 and 200~GeV, an agreement is observed within uncertainties. Among the three models (UrQMD, HIJING, AMPT) compared, UrQMD provides the best description of the experimental data, particularly at lower energies and in non-central collisions where the spectator contribution is significant. However, it overestimates the yield at $\sqrt{s_{_{\text{NN}}}} = $ 200~GeV.

\section*{Acknowledgements}

We thank the RHIC Operations Group and SCDF at BNL, the NERSC Center at LBNL, and the Open Science Grid consortium for providing resources and support.  This work was supported in part by the Office of Nuclear Physics within the U.S. DOE Office of Science, the U.S. National Science Foundation, National Natural Science Foundation of China, Chinese Academy of Science, the Ministry of Science and Technology of China and the Chinese Ministry of Education, NSTC Taipei, the National Research Foundation of Korea, Czech Science Foundation and Ministry of Education, Youth and Sports of the Czech Republic, Hungarian National Research, Development and Innovation Office, New National Excellency Programme of the Hungarian Ministry of Human Capacities, Department of Atomic Energy and Department of Science and Technology of the Government of India, the National Science Centre and WUT ID-UB of Poland, German Bundesministerium f\"ur Bildung, Wissenschaft, Forschung and Technologie (BMBF), Helmholtz Association, Ministry of Education, Culture, Sports, Science, and Technology (MEXT), Japan Society for the Promotion of Science (JSPS), and Agencia Nacional de Investigacion y Desarrollo de Chile (ANID), Chile.

\printbibliography

\end{multicols}

\end{document}